\documentclass[aps,prl,twocolumn,amsmath,amssymb,nofootinbib,superscriptaddress,floatfix]{revtex4}
\usepackage{amsmath}
\usepackage{amssymb}
\usepackage{amsthm}
\usepackage[pdftex,colorlinks,citecolor=blue,linkcolor=red,urlcolor=blue]{hyperref} 
\usepackage{graphicx}
\usepackage{dcolumn} 
\usepackage{bm} 
\usepackage{longtable}
\usepackage{ulem}   
\usepackage{comment} 
\usepackage{datetime}

\newcommand{\bs}[1]{{\boldsymbol{#1}}}

\newcommand{\be}{\begin{equation}}
\newcommand{\ee}{\end{equation}}
\newcommand{\bea}{\begin{eqnarray}}
\newcommand{\eea}{\end{eqnarray}}
\newcommand{\barr}{\begin{array}}
\newcommand{\earr}{\end{array}}

\long\def\begincomment#1\endcomment{}

\newcommand{\cH}{\mathcal{H}}

\newcommand{\GSD}{\mathop{\mathrm{GSD}}}
\newcommand{\tr}{\mathop{\mathrm{tr}}}

\begin{document}

\title{ 
Symmetry-protected many-body Aharonov-Bohm effect
      }

\author{Luiz H. Santos} \email{lsantos@perimeterinstitute.ca}
\affiliation{
Perimeter Institute for Theoretical Physics,
Waterloo, Ontario, N2L 2Y5, Canada
            }

\author{Juven C. Wang}  \email{juven@mit.edu}
\affiliation{
Department of Physics, Massachusetts Institute of Technology,
Cambridge, MA 02139, USA
            } 
\affiliation{
Perimeter Institute for Theoretical Physics,
Waterloo, Ontario, N2L 2Y5, Canada
            }

\begin{abstract}
It is known as a purely quantum effect that a magnetic flux affects the real physics of a particle, such as the energy spectrum, even if the flux does not interfere with the particle's path - the Aharonov-Bohm effect. Here we examine an Aharonov-Bohm effect on a many-body wavefunction. Specifically, we study this many-body effect on the gapless edge states of a bulk gapped phase protected by a global symmetry (such as $\mathbb{Z}_{N}$) - the symmetry-protected topological (SPT) states. 
The many-body analogue of spectral  shifts, the twisted wavefunction and the twisted boundary realization are identified in this SPT state. 
An  explicit lattice construction of SPT edge states is derived, and a challenge of gauging its non-onsite symmetry is overcome.  Agreement is found in the twisted
spectrum between a numerical lattice calculation and a conformal field theory prediction.
\end{abstract}

\maketitle

\begin{figure}[h!]
\includegraphics[width=0.5\textwidth]{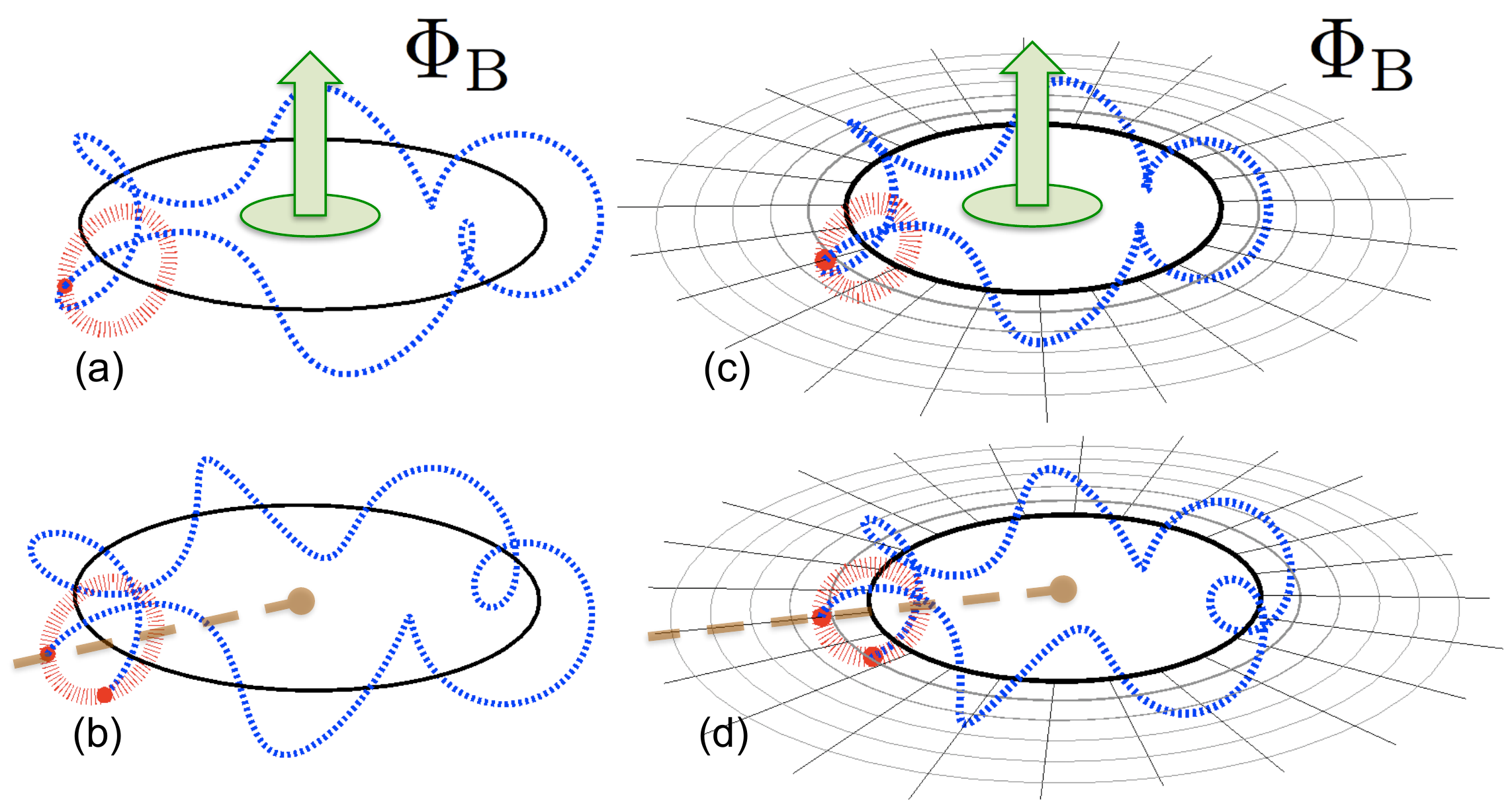}
\caption{
(a) and (c): Single- and many-body wavefunctions upon
flux insertion, respectively.
(b) and (d): Flux effect captured by twisted boundary conditions
showing the associated branch cut.
}
\label{fig:Ring/Edge geometries and wavefunction twist}
\end{figure}

Mysteriously
an external magnetic flux can  affect the physical properties of particles 
even without interfering directly on their paths.
It is known as the Aharonov-Bohm(AB) effect~\cite{Aharonov-Bohm}.
For instance, a particle of charge $q$ and mass $\mathfrak{m}$ confined in a ring 
(parametrized by 
$
0 \leq \theta < 2\pi
$)
of radius 
$\mathfrak{a}$ 
threaded with a flux $\Phi_{B}$,
see Fig.~\ref{fig:Ring/Edge geometries and wavefunction twist}(a),
would have its energy spectrum 
shifted as
\begin{equation}
E_{n}
=
\frac{1}{2\mathfrak{m}\mathfrak{a}^2}\,
\left(
n
+
\frac
{
\Phi_{B}
}
{
\Phi_{0}
}
\right)^2,
\quad
n = 0, \pm 1, ...,
\label{eq:particle on a ring spectrum}
\end{equation}  
where $\Phi_{0} = 2\pi/q$
is the quantum of magnetic flux and 
we adopt $e=\hbar = c = 1$ units.
One can dispose of the gauge potential in the Schr\"odinger\rq{s} equation
of the wavefunction
$
\psi(\theta)
$,
by a gauge transformation that changes the wavefunction to
$
\tilde{\psi}(\theta)
= 
\psi(\theta)\,\exp[ i\,q \int^{\theta}\bs{A}(\theta\rq{})\,d\,\theta\rq]
$.
So, the effect of the external flux can be 
enforced by the condition that the phase 
$
\tilde{\varphi}(\theta)
$
of the 
new wavefunction satisfies 
a twisted boundary condition,
\begin{equation}
(1/2\pi)\,
\oint\,d\theta\,
\frac
{
\partial\,\tilde{\varphi}(\theta)
}
{
\partial \theta
}
=
\Phi_{B}/\Phi_{0},
\label{eq:twisted BC for single particle state}
\end{equation}
as the particle trajectory encloses the ring;
thus, this twisted boundary condition
implies a \lq\lq{branch cut}\rq\rq,
see Fig.~\ref{fig:Ring/Edge geometries and wavefunction twist}(b).
We may refer to this twist effect as \lq\lq{{\it Aharonov-Bohm twist}}\rq\rq.
For electrons confined on a mesoscopic ring, for
example, even though interactions are not negligible,
the sensitivity of the system to the presence of the external flux
can be rationalized as a single particle phenomenon~\cite{Aronov-Sharvin}.

It is then opportune, as matter of principle,
to ask whether such an AB effect can
take place as an \textit{intrinsically} interacting
many-body phenomenon.
More concretely, we ask whether the low energy
properties of such interacting systems display a response analogous
to Eq.~(\ref{eq:particle on a ring spectrum}) when subject to a gauge 
perturbation and, in turn, how this effect is encoded
in the ``topology\rq{}\rq{} (or boundary conditions) 
of the the wave-functional 
$
\Psi[\phi(x)]
$,
see Figs.~\ref{fig:Ring/Edge geometries and wavefunction twist}(c)-(d). 
We shall refer to this as 
a many-body AB effect or twist.

In this paper we show that $2$D \lq\lq{symmetry protected topological}\rq\rq (SPT) 
states~[\onlinecite{Chen:2011pg,2011PhRvB..84w5141C,Wen:2013ue}] 
offer a natural platform for observing the many-body AB effect.
SPT states are quantum many-body states of matter with a finite gap to bulk excitations and no fractionalized degrees of freedom. 
Due to a global symmetry, the system has the property that its edge states can 
only be gapped if a symmetry breaking occurs, either explicitly or spontaneously.
So, in the absence of any symmetry breaking, the edge is described 
by robust edge excitations which can not be localized due to weak symmetry-preserving disorder, in contrast to purely one dimensional systems~\cite{Lee-Ramakrishnan}. Assuming then that the edge states are in this gapless
phase (an assumption which we will take throughout the paper), we shall demonstrate that the system will respond to the insertion of a gauge flux in a 
non-trivial way, whereas if the edge degrees of freedom were to become gapped, then they would be insensitive to the flux.
We note that in 2D systems displaying the integer quantum Hall effect, the insertion of a flux also
induces a non-trivial response of the chiral edge states~\cite{Laughlin1981}.
In contrast to this situation, here we shall be concerned with 2D non-chiral SPT 
states for which the gapless edge excitations, like the single 
particle modes on a ring, propagate in both directions. The spectrum of these
gapless modes characterize the low energy properties of the system.

We approach this problem from two venues: 
(I) First we study the response of the SPT state
to the insertion of a gauge flux by means of a low energy effective theory
for the edge states
and derive the change in the 
spectrum of edge states akin to Eq.~(\ref{eq:particle on a ring spectrum}).
(II) Complementarily, we show that the many-body AB effect derived
in (I) can also be captured by formulating a lattice model
describing the edge states. Twisted boundary conditions 
defined for these models are shown to account for the
presence of a gauge flux, which we confirm numerically.

\section{Many-body Aharonov-Bohm effect}

To capture the essence of AB effect on a symmetry-protected many-body wavefunction,  we imagine threading a gauge flux through an effective 1D edge 
on one side of a $2$D bulk SPT annulus (or cylinder). This many-body wavefunction on the $1$D edge
(parametrized by $0 \leq x <L$)
of SPT states is the analogue of a single-body wavefunction of a particle in a ring.
Since the bulk degrees of freedom are gapped,
we concentrate on the low energy properties on the edge
described by a non-chiral Luttinger liquid action
$
I_{\textrm{edge}}[\phi_{I}]
$
[\onlinecite{LevinGu,Lu:2012dt}]. 
To capture the gauge flux effect on a many-body wave function $| \Psi \rangle$, 
we formulate it in the path integral,
\begin{equation}
\begin{split}
&\,
| \Psi(t_f) \rangle 
=
\sum_n | \Psi_n(t_f) \rangle  \langle \Psi_n(t_f)   | e^{-i \int^{t_f}_{t_i} H(t)dt}| \Psi(t_i) \rangle   
\\
&\,
=
\sum_n | \Psi_n(t_f) \rangle  \int^{\phi_{I,n}}_{\phi_{I}(t_i)} \mathcal{D}\phi_{I}
e^
{
i\,(I_{\textrm{edge}}[\phi_{I}]+
(1/2\pi) \int q^{I} \bs{A} \wedge d \phi_I)
},
\end{split}  
\label{eq:gauged edge theory}
\end{equation}
with $\phi_I$ the intrinsic field on the edge.
Our goal 
is to interpret this many-body AB twist 
$(1/2\pi) \int q^{I} \bs{A} \wedge d \phi_I$.
We anticipate the energy spectrum under the flux would be adjusted, and we aim to capture this ``twist'' effect on the energy spectrum.
Below we 
focus on bosonic SPT states with $\mathbb{Z}_N$ symmetry
[\onlinecite{LevinGu,Lu:2012dt,Chenggu,Hung:2013nla,Ye:2013upa}],
with global symmetry transformation on the edge
(see
Appendix 1
for details on the field theoretic input)
\begin{equation}
\mathcal{S}^{(p)}_{N}
=
e^
{
\frac{i}{N}\,
\left(
\int^{L}_{0}\,dx\,\partial_{x}\phi_{2}
+
p\,\int^{L}_{0}\,dx\,\partial_{x}\phi_{1}
\right)
},
\label{eq:SPT symmetry transformation phi_1 and phi_2}
\end{equation}  
where
$
p \in \{ 0, ..., N-1 \}
$
and
$
(1/2\pi)\partial_{x}\phi_{2}(x)
$
is the canonical momentum associated to
$
\phi_{1}(x)
$ [\onlinecite{Wang:2012am}].

The Lagrangian density associated to Eq.(\ref{eq:gauged edge theory}) reads
\begin{equation}
\mathcal{L}_{\textrm{edge}}[A]
=
\frac{1}{4\pi}
K_{IJ}\,
\partial_{t}\phi_{I}\,
\partial_{x}\phi_{J}\,
-
\mathcal{H}_{f}[\phi_{I}]
+
\frac{1}{2\pi}
q^{I}\,A_{\mu}\varepsilon^{\mu\nu}\partial_{\nu}\phi_{I},
\label{eq:edge lagrangian with A}
\end{equation}  
where indices $\mu,\nu\in\{0,1\}$, $I,J\in\{1,2\}$,
$
K
=
\begin{pmatrix}
0 & 1
\\
1 & 0
\end{pmatrix}  
$,
$
\mathcal{H}_{f}[\phi_{I}]
$
is the Hamiltonian density describing a 
free boson
and
$
q^{I} = (q^{1},q^{2}) = (1,p)
$
specify the charges 
{
carried by the currents
$
J^{\mu}_{I}
=
(1/2\pi)
\varepsilon^{\mu\nu}
\partial_{\nu}\phi_{I}
$.
}
The right/left moving modes are described by
$
\phi_{R,L}
\propto
\phi_{1}
\pm
\phi_{2}
$.

Integrating the equations of motion of~(\ref{eq:edge lagrangian with A}),
with respect to $\phi_{I}$, along the boundary coordinate $x$
in the presence of a static 
background
$\mathbb{Z}_{N}$
gauge flux configuration
$
\oint^{L}_{0} dx A_{1}(x)
=
\frac{2\pi}{N}
$
delivers
\begin{equation}
\begin{split}
&\,
(1/2\pi)\oint^{L}_{0} dx\,\partial_{x}
\begin{pmatrix}
\phi_{1} 
\\
\phi_{2}
\end{pmatrix} 
=
\frac{1}{N}
\begin{pmatrix}
p
\\
1
\end{pmatrix}.
\end{split}  
\label{eq:change in winding modes}
\end{equation}
(See 
Appendix 2
for an alternative derivation from a bulk-edge Chern-Simons approach).
Eq.~(\ref{eq:change in winding modes})
represents the shift in winding modes of the edge boson fields
and plays a role analogous to the single-particle twisted
boundary condition Eq.~(\ref{eq:twisted BC for single particle state}).
The spectrum of the central charge $c = 1$
free boson at compactification radius $R$
is labeled by the primary states $|n,m\rangle$ ($n,m \in \mathbb{Z}$)
with scaling dimension
\begin{equation}
\Delta(n,m;R)
=
\frac{n^2}{R^2}
+
\frac{R^2 m^{2}}{4}
\label{eq:scaling dimensions ungauged case}
\end{equation}
and momentum
$
\mathcal{P}(n,m) = n m
$~\cite{DiFrancesco}.
Then, according to Eq.~(\ref{eq:change in winding modes}), 
after the flux insertion, we derive the new spectrum (also see another related setting~\cite{Sule:2013qla}) 
\begin{equation}
\begin{split}
&\,
\tilde{\Delta}^{(p)}_{N}(n,m;R)
=
\frac{1}{R^2}\left(n+\frac{p}{N}\right)^{2}
+
\frac{R^2}{4}\left(m + \frac{1}{N}\right)^{2}
\end{split}  
\label{eq:scaling dimensions with flux}
\end{equation}  
and momenta
$
\tilde{\mathcal{P}}^{(p)}_{N}(n,m)
=
\left(n +\frac{p}{N} \right) \left(m +\frac{1}{N} \right)
$
for each SPT state 
$
p \in \{ 0, ..., N-1 \}
$.
Eqs.~(\ref{eq:change in winding modes})
and~(\ref{eq:scaling dimensions with flux})
capture the essence of the many-body AB effect
analogous to Eqs.~(\ref{eq:particle on a ring spectrum})
and~(\ref{eq:twisted BC for single particle state}).

\section
{
Effective lattice model for the
edge of SPT states
}

\noindent
{\it -Symmetry transformation and domain wall-}  \\
The twist effect encoded in Eq.~(\ref{eq:scaling dimensions with flux})
comes from an effective low energy description of the edge. 
We aim, as a complementary and perhaps more fundamental point of view,
to capture this twist effect from a lattice model.
As a first step in this program, we shall construct a global $\mathbb{Z}_{N}$
symmetry transformation in terms of discrete degrees of freedom on the edge
whose action reduces to Eq.~(\ref{eq:SPT symmetry transformation phi_1 and phi_2})
at long wavelengths. The hallmark of a non-trivial SPT state is that
the symmetry transformation  
on the boundary cannot be in a tensor product form on each single 
site, i.e., it acts as a non-onsite symmetry transformation [\onlinecite{Chen:2011pg,2011PhRvB..84w5141C,Chen:2012hc}].
We propose the following ansatz for the symmetry transformation,
\begin{equation}
\begin{split}
\bullet\;\;\; 
S^{(p)}_{N}
&\,\equiv
\prod^{M}_{j=1} 
\tau_j  
\prod^{M}_{j=1} 
\exp\Big\{ i \frac{p}{N}  \big[\frac{2\pi}{N}(\delta N_{\text{DW}})_{j,j+1} \big]\Big\}
\\
&\,
\equiv 
\prod^{M}_{j=1} \tau_j \;
\prod^{M}_{j=1}  
e^
{
\frac{i}{N}\,Q^{(p)}_{N}(\sigma^{\dagger}_{j}\sigma_{j+1})
}, 
\end{split}
\label{eq:symmetry form lattice} 
\end{equation}
acting on a ring with $M$ sites that we take to describe
the $1$D edge,
with $\sigma_{M+1}\equiv\sigma_{1}$.
At every site of the ring we consider a pair of two $\mathbb{Z}_{N}$ operators:
$(\tau_{j},\sigma_{j})$, with a site index $j = 1,...,M$, satisfying
$\tau^{N}_{j} = \sigma^{N}_{j} = \openone$ and 
a conjugation relation $\tau^{\dagger}_{j}\,\sigma_{j}\,\tau_{j} = \omega\,\sigma_{j}$, where $\omega \equiv e^{i\,2\pi/N}$.
We shall use the following representation 
\begin{equation}
\begin{split}
&\,
\sigma_{j} 
=
{\begin{pmatrix} 
1 & 0 & 0 & 0 \\
0 & \omega & 0 & 0\\  
0 & 0 & \ddots  & 0\\  
0 & 0 & 0 & \omega^{N-1}
\end{pmatrix}}_j\,,
%
\\
&\,
\tau_{j} 
=
{\begin{pmatrix} 
0 & 0 & 0 & \dots &0& 1  \\
1 & 0 & 0 & \dots &0& 0 \\
0 & 1 & 0 & \dots &0& 0 \\
0 & 0 & 1 & \dots &0& 0 \\
\vdots &0 & 0 & \dots &1 & 0 
\end{pmatrix}}_j
\,.
\end{split}
\end{equation}
The operators act on the Hilbert space of $\mathbb{Z}_{N}$ states for each site $j$.

The overall symmetry transformation contains 
the onsite transformation part 
generated by the string of $\tau$\rq{}s and
the ``non-onsite {\it domain wall} (DW)'' part
$
(
\delta N_{\text{DW}}
)_{j,j+1}
$
between sites $j$ and $(j+1)$.
The ansatz form Eq.~(\ref{eq:symmetry form lattice})
has the property that 
$
\prod^{M}_{j=1}\tau_{j}
$
and
$
\prod^{M}_{j=1}
e^
{
\frac{i}{N}\,Q^{(p)}_{N}(\sigma^{\dagger}_{j}\sigma_{j+1})
}
$
commute,
and
the unitarity of $S^{(p)}_{N}$
implies
$
[ Q^{(p)}_{N} ]^{\dagger} = Q^{(p)}_{N}
$.
It follows that 
\begin{equation}
\Big(\,S^{(p)}_{N}\,\Big)^N
=
\prod^{M}_{j=1}\,
e^
{
i\,Q^{(p)}_{N}(\sigma^{\dagger}_{j}\sigma_{j+1})
}\,.
\end{equation}•
The construction above then naturally yields $N$ distinct classes of 
$\mathbb{Z}_{N}$ symmetry transformations, labeled by $p \in \mathbb{Z}_{N}$,
upon imposing
the following condition 
on the $(N-1)$-th order polynomial
operator 
$
Q^{(p)}_{N}(\sigma^{\dagger}_{j}\sigma_{j+1})$:
\begin{equation}
e^
{
i\,Q^{(p)}_{N}(\sigma^{\dagger}_{j}\sigma_{j+1})
}
=
(\sigma^{\dagger}_{j}\sigma_{j+1})^{p},
\quad
p = 0,...,N-1,
\label{eq:Z_n condition on $U_{j,j+1}$}
\end{equation}  
which guarantees (due to periodic boundary conditions)
that 
$
( S^{(p)}_{N} )^{N} = \openone
$.
The symmetry transformation in the trivial case corresponds to
$
p=0\, (\text{mod}\,N)
$
for which
$
\prod_{j}
e^
{
\frac{i}{N}\,Q^{(p=0)}_{N}(\sigma^{\dagger}_{j}\sigma_{j+1})
}
=
\openone
$,
while
$
p
\neq
0
\, (\text{mod}\,N)$
describe the other $N-1$ non-trivial SPT classes.
Identifying
$
\sigma_{j}
\sim
e^
{
i\,\phi_{1}(j)
}
$,
then
the domain wall variable $(\delta N_{\text{DW}})_{j,j+1}$ counts the number of units of $\mathbb{Z}_N$ angle between sites $j$ and $j+1$, so
$(2\pi/N)(\delta N_{\text{DW}})_{j,j+1}$$=\phi_{1,j+1} - \phi_{1,j}$,
which produces the expected long distance behavior
of the symmetry transformation Eq.~(\ref{eq:SPT symmetry transformation phi_1 and phi_2}).
Our ansatz nicely embodies two interpretations together, on both a continuum field theory and a discrete lattice model.
The $\mathbb{Z}_{N}$ symmetry transformations Eq.~(\ref{eq:symmetry form lattice})
that satisfy Eq.~(\ref{eq:Z_n condition on $U_{j,j+1}$}) can be explicitly written as
\begin{equation}
S^{(p)}_{N}
=
\prod^{M}_{j=1} 
\tau_j  
\,
\prod^{M}_{j=1}  
e^
{
-i\frac{2\pi}{N^2}p\, 
\Big\{
\left(
\frac{N-1}{2}\,
\right)
\openone
+
\sum^{N-1}_{k=1}\,
\frac
{
(\sigma^{\dagger}_{j}\sigma_{j+1})^k
}
{
(\omega^k - 1)
}
\Big\}
}. 
\label{eq:symmetry form lattice 2}
\end{equation}•
%
In Ref.\onlinecite{Chen:2012hc} the edge symmetry for $\mathbb{Z}_{N}$
SPT states was proposed in terms of effective long-wavelength rotor variables.
We emphasize
that the construction of the edge symmetry 
transformations Eq.~(\ref{eq:symmetry form lattice 2})
described here does not rely on a long wavelength description; rather it can be viewed
as a fully regularized symmetry transformation.
In Appendices 3 and 4, we give explicit formulas for the 
$
\mathbb{Z}_{2}
$
and
$
\mathbb{Z}_{3}
$
symmetry transformations,
as well as we draw a connection between
the lattice operators $(\tau_{j},\sigma_{j})$
and quantum rotor variables.
\\

\noindent
{\it -Lattice model-}  \\ 
Having constructed all the classes of
$\mathbb{Z}_{N}$ symmetry transformations 
Eq.~(\ref{eq:symmetry form lattice 2}),
we now propose our 
translation invariant and $\mathbb{Z}_{N}$-symmetric
lattice model Hamiltonians $H^{(p)}_{N}$
on the edge of $\mathbb{Z}_{N}$ SPT states, i.e.,
\be \label{principle1-2}
[H^{(p)}_{N},T]=0, \;\;  
[H^{(p)}_{N},
S^{(p)}_{N}
]
=0,
\ee
where
$
T
$
performs a translation by one lattice site.
Our model Hamiltonian is (with $\lambda^{(p)}_{N}$ a constant),
\begin{equation}
H^{(p)}_{N}
\!\!
=
\!
\lambda^{(p)}_{N}
\!
\sum^{M}_{j=1}
\!
h^{(p)}_{N,j}
\!
\equiv
\!
-
\lambda^{(p)}_{N}
\!
\sum^{M}_{j=1} 
\!\!
\sum^{N-1}_{\ell = 0}
\!\!
\left(
S^{(p)}_{N} 
\right)^{\!\!-\ell}
\!\!\!
( \tau_{j} + \tau^{\dagger}_{j} )  
\!\!
\left(
S^{(p)}_{N} 
\right)^{\ell}.
\label{eq:Hamiltonian lattice}
\end{equation}
Notice that $H^{(p)}_{N}$ is manifestly
$\mathbb{Z}_{N}$ symmetric since
it is constructed from the superposition of $\tau_{j}$ conjugated to all
powers of $S^{(p)}_{N}$.
In the trivial SPT case for which
$
H^{(p=0)}_{N}
\propto
-
\sum^{M}_{j=1}\,
( \tau_{j} + \tau^{\dagger}_{j} )  
$,
the model gives a gapped and symmetry preserving ground state.
In Appendix 3, we provide
explicit forms of the non-trivial classes of SPT Hamiltonians
for the $N=2$ and $N=3$ cases. 
We note that for the $\mathbb{Z}_{2}$ case, our symmetry transformation
and edge Hamiltonian are the same as that obtained in Ref. \onlinecite{LevinGu}
(where the low energy theory in terms of a non-chiral Luttinger liquid has been discussed),
despite the fact that our method of constructing the symmetry is independent of that in Ref. \onlinecite{LevinGu} and provides a generalization for 
all $\mathbb{Z}_{N}$ groups. It is noteworthy to mention that the authors of 
Ref. \onlinecite{LevinGu} argue that the edge of the $\mathbb{Z}_{2}$ bosonic SPT state is
generically unstable to symmetry preserving perturbations. Nevertheless, we shall still 
study the model Hamiltonian~(\ref{eq:Hamiltonian lattice}) for the $\mathbb{Z}_{2}$ as a means to address 
our numerical methods. 
A common feature of these Hamiltonian classes
is the existence of combinations of terms like 
$\sigma_{j-1}\tau_{j}\sigma_{j+1}$ due to the non-onsite global symmetry.
Their effect, as we shall see, is to give rise to a gapless spectrum.
In order to understand their effect on the low energy properties,
we perform an exact diagonalization study of the non-trivial 
Hamiltonian classes Eq.~(\ref{eq:Hamiltonian lattice}) 
on finite systems.

\begin{figure}[h!]
\includegraphics[width=0.5\textwidth]{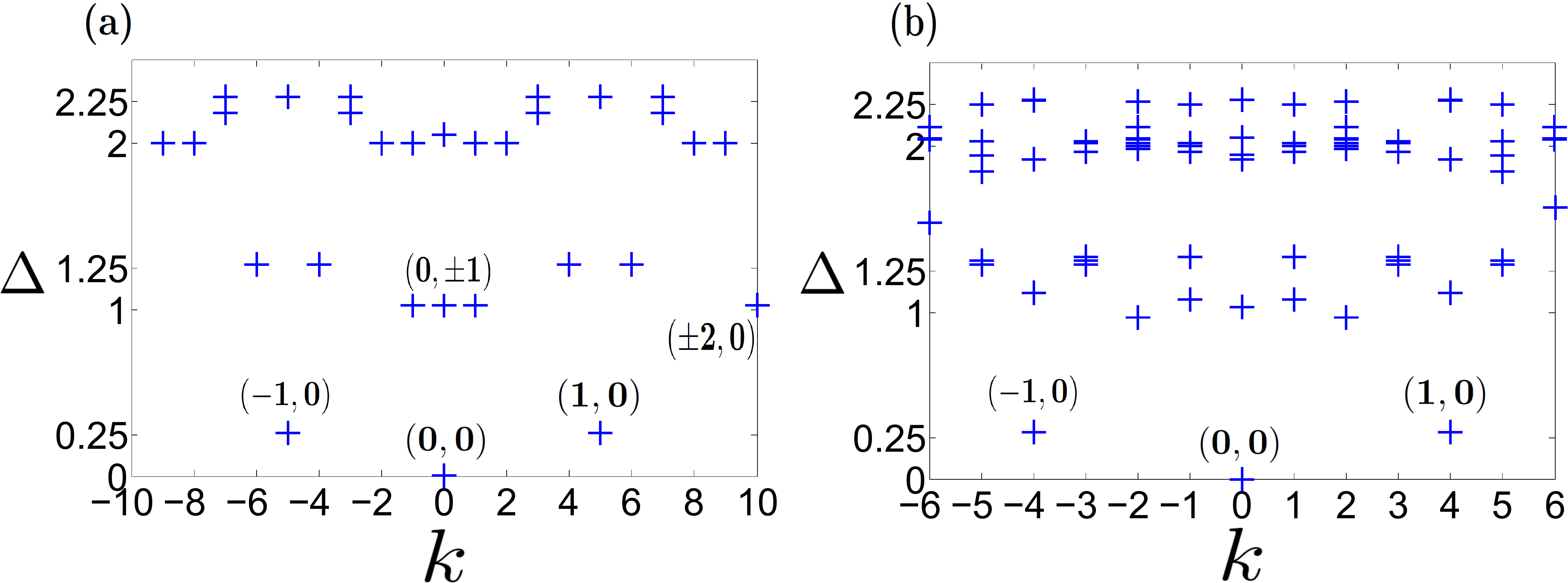}
\caption
{
Spectrum of the SPT Hamiltonian Eq.~(\ref{eq:Hamiltonian lattice})
with respect to the lowest energy $E^{(p)}_{N,0}$,
on a ring as a function of the lattice momentum
$k \in \mathbb{Z}$.
First few primary states labeled by $(n,m)$.
(a) Spectrum of $H^{(p=1)}_{2}$ with $\lambda^{(p=1)}_{2} = 0.82$ 
and $M = 20$ sites.
(b) Spectrum of $H^{(p=1,2)}_{3}$ with $\lambda^{(p=1,2)}_{3} =0.26$ 
and $M = 12$ sites.
The values of $\lambda^{(p)}_{N}$ above guarantee a proper normalization so that
states in the same conformal tower separated by $\delta k  = \pm 1$
are integer spaced (up to finite size effects) [see Ref.~\onlinecite{Henkel}].
}
\label{fig:spectrum untwisted SPT}
\end{figure}

In Fig.~(\ref{fig:spectrum untwisted SPT}) we 
plot
the
lowest energy eigenvalues for the 
$\mathbb{Z}_{2}$ and $\mathbb{Z}_{3}$ non-trivial SPT states as 
a function of the lattice momentum $k \in \mathbb{Z}$ defined
by
$
T
=
e^
{
i\frac{2\pi}{M}k
}
$.
The spectrum of $H^{(1)}_{2}$ with $M=20$ sites,
shows very good agreement with 
the bosonic spectrum Eq.~(\ref{eq:scaling dimensions ungauged case})
at
$
R
=
2
$,
with states being labeled by $|n,m\rangle$.
The global $\mathbb{Z}_{2}$ charges 
relative to the ground state 
were found to be 
$
e^
{
i\,\pi\,(n + m)
}
$
in accordance to Eq.~(\ref{eq:SPT symmetry transformation phi_1 and phi_2})
(We note that similar results have been obtained for the
$\mathbb{Z}_{2}$ case in Ref.~\onlinecite{Chen:2012hc}).
For the $\mathbb{Z}_{3}$ SPT states,
which have not been investigated before, with $M=12$ sites, the 
spectrum of $H^{(1)}_{3}$ and $H^{(2)}_{3}$ are identical~\cite{H}. 
Finite size effects are more prominent than in the $\mathbb{Z}_{2}$ case, 
but the overall structure of the spectrum is very similar with the second and 
third states being degenerate with energy close to $1/4$ and
global 
$\mathbb{Z}_{3}$
charges
$
e^
{
\pm 2\pi i / 3
}
$
(which we identify as the $|n=\pm1,m=0\rangle$ states),
suggesting the same spectrum Eq.~(\ref{eq:scaling dimensions ungauged case}) at $R=2$.

In Appendix 4, following the methods of Refs.~[\onlinecite{Chen:2012hc,2011PhRvB..84w5141C,Chen:2011pg}],
we show that the symmetry classes defined in Eq.~(\ref{eq:symmetry form lattice}) subject to
condition Eq.~(\ref{eq:Z_n condition on $U_{j,j+1}$}) are related to all 
$\mathbb{Z}_{N}$ $3$-cocycles of the group cohomology classification of $2$D SPT states [\onlinecite{Chen:2011pg}].
Thus, our lattice model completely realizes all $N$ classes of $\mathcal{H}^{3}(\mathbb{Z}_N,U(1))=\mathbb{Z}_N$, 
where
$p$ stands for the $p$-th class in the third cohomology group.

\section{Twisted boundary conditions and Twisted Hamiltonian on the lattice}

We now seek to 
build a lattice model with twisted boundary conditions
to capture the edge states spectral shift
in the presence of a unit of 
$\mathbb{Z}_{N}$ flux insertion. 

It is instructive to revisit the case of twisted boundary conditions
where the symmetry transformation acts as an on-site symmetry.
For the sake of concreteness, let us consider the one dimensional quantum Ising model
$
H_{Ising}
=
\sum^{M}_{j=1}\,(J\,\sigma^{z}_{j}\,\sigma^{z}_{j+1} +h\, \sigma^{x}_{j} )
$
with global $\mathbb{Z}_{2}$ symmetry
$
\prod^{M}_{j=1}\,\sigma^{x}_{j}
$.
The $\mathbb{Z}_{2}$ twisted sector 
(or equivalently, in this case, the anti-periodic boundary condition sector)
of the model
is realized by flipping the sign of a pair interaction 
$
\sigma^{z}_{k}\sigma^{z}_{k+1}
\rightarrow
-
\sigma^{z}_{k}\sigma^{z}_{k+1}
$,
for some site $k$,
while leaving all the other terms unchanged. If the Ising model
is defined on an open line, the twist effect is implemented by
conjugating the $H_{Ising}$ with the operator
$
\prod_{\ell \leq k}\,\sigma^{x}_{\ell}.
$
When the model is defined on a ring, the same effect 
is obtained by defining a new translation operator
$\tilde{T} = T\,\sigma^{x}_{k}$ and demanding 
that the twisted Hamiltonian $\tilde{H}_{Ising}$ commutes with $\tilde{T}$.
It is straightforward to see that the twisted Ising Hamiltonian on a ring which commutes with
$\tilde{T}$ indeed has
$
\sigma^{z}_{k}\sigma^{z}_{k+1}
\rightarrow
-
\sigma^{z}_{k}\sigma^{z}_{k+1}
$.
We also note that 
$
(\tilde{T})^M = \prod^{M}_{j=1}\,\sigma^{x}_{j}
$
generates the $\mathbb{Z}_{2}$ symmetry of $H_{Ising}$,
which is also a symmetry of $\tilde{H}_{Ising}$.

We now generalize the construction above for the
SPT edge Hamiltonians on a ring with a non-onsite symmetry by 
defining the unitary twisted lattice translation operator\cite{many-flux-case}
\begin{equation}
\tilde{T}^{(p)}
=
T\,
e^
{
\frac{i}{N}\,
Q^{(p)}_{N}
(
\sigma^{\dagger}_{M}\sigma_{1}
)
}
\tau_{1},
\end{equation}  
for each 
$p \in \mathbb{Z}_N$
classes, which incorporates the effect of the branch cut 
as in Fig.~\ref{fig:Ring/Edge geometries and wavefunction twist}(d).
The twisted Hamiltonian 
$
\tilde{H}^{(p)}_{N}
$,
constructed from
$
{H}^{(p)}_{N}
$
of Eq.~(\ref{eq:Hamiltonian lattice})
and satisfying
\begin{equation} 
[ \tilde{H}^{(p)}_{N}, \tilde{T}^{(p)} ] =0,\;\;\; 
\label{principle2-2}
\end{equation}
reads (see Appendix 3.2 for explicit results)
\begin{subequations}
\begin{equation}
\tilde{H}^{(p)}_{N}
=
\lambda^{(p)}_{N}
\sum^{M}_{j=1}
\tilde{h}^{(p)}_{N,j}
\end{equation}•
\begin{equation}
\begin{split}
&\,
\tilde{h}^{(p)}_{N,1}
=
\tau^{\dagger}_{1}\,\tau^{\dagger}_{2}\,
h^{(p)}_{N,1}
\tau_{1}\,\tau_{2},
\\
&\,
\tilde{h}^{(p)}_{N,j}
=
h^{(p)}_{N,j} ~ (2  \leq j \leq M-1),
\\
&\,
\tilde{h}^{(p)}_{N,M}
=
\tau^{\dagger}_{1}\,
e^
{
-\frac{i}{N}\,
Q^{(p)}_{N}
(
\sigma^{\dagger}_{M}\sigma_{1}
)
}
\,
h^{(p)}_{N,M}
\,
e^
{
\frac{i}{N}\,
Q^{(p)}_{N}
(
\sigma^{\dagger}_{M}\sigma_{1}
)
}
\tau_{1}.
\end{split}  
\end{equation}
\label{eq:Hamiltonian lattice-twisted a}
\end{subequations}•

Notice that, due to the intrinsic non-onsite term in the
symmetry transformation,
\begin{equation} 
\tilde{S}^{(p)}_{N} \equiv
( \tilde{T}^{(p)} )^{M}
=
e^
{
\frac{i}{N}
\Big[
Q^{(p)}_{N}(\omega\,\sigma^{\dagger}_{M}\,\sigma_{1})
-
Q^{(p)}_{N}(\sigma^{\dagger}_{M}\,\sigma_{1})
\Big]
}
\,
S^{(p)}_{N}\,,
\label{principle2-1 a}
\end{equation}
the twisted non-trivial
Hamiltonian breaks the SPT global symmetry (i.e. 
$
[\tilde{H}^{(p)}_{N}, S^{(p)}_{N}  ]
\neq
0
$
if
$p
\neq
0
$
mod($N$)\,),
signaling an anomaly effect\cite{Wen:2013oza,Wang:2014tia}.
(For a more systematic discussion of bosonic anomalies 
in the context of $2$D SPT states, see Ref.[\onlinecite{Wang:2014tia}].)
However, in the trivial state, 
Eq.~(\ref{principle2-1 a}) yields
$
\tilde{S}^{(p=0)}_{N} 
=
S^{(p=0)}_{N}
=
\prod^{M}_{j=1}\tau_{j}
$,
so that the twisted trivial Hamiltonian still \textit{commutes}
with the global $\mathbb{Z}_{N}$ onsite symmetry,
and the twisted effect is equivalent to usual toroidal 
boundary conditions~\cite{Henkel},
as exemplified before for anti-periodic boundary condition of the Ising model.

\begin{figure}[h!]
\includegraphics[width=0.5\textwidth]{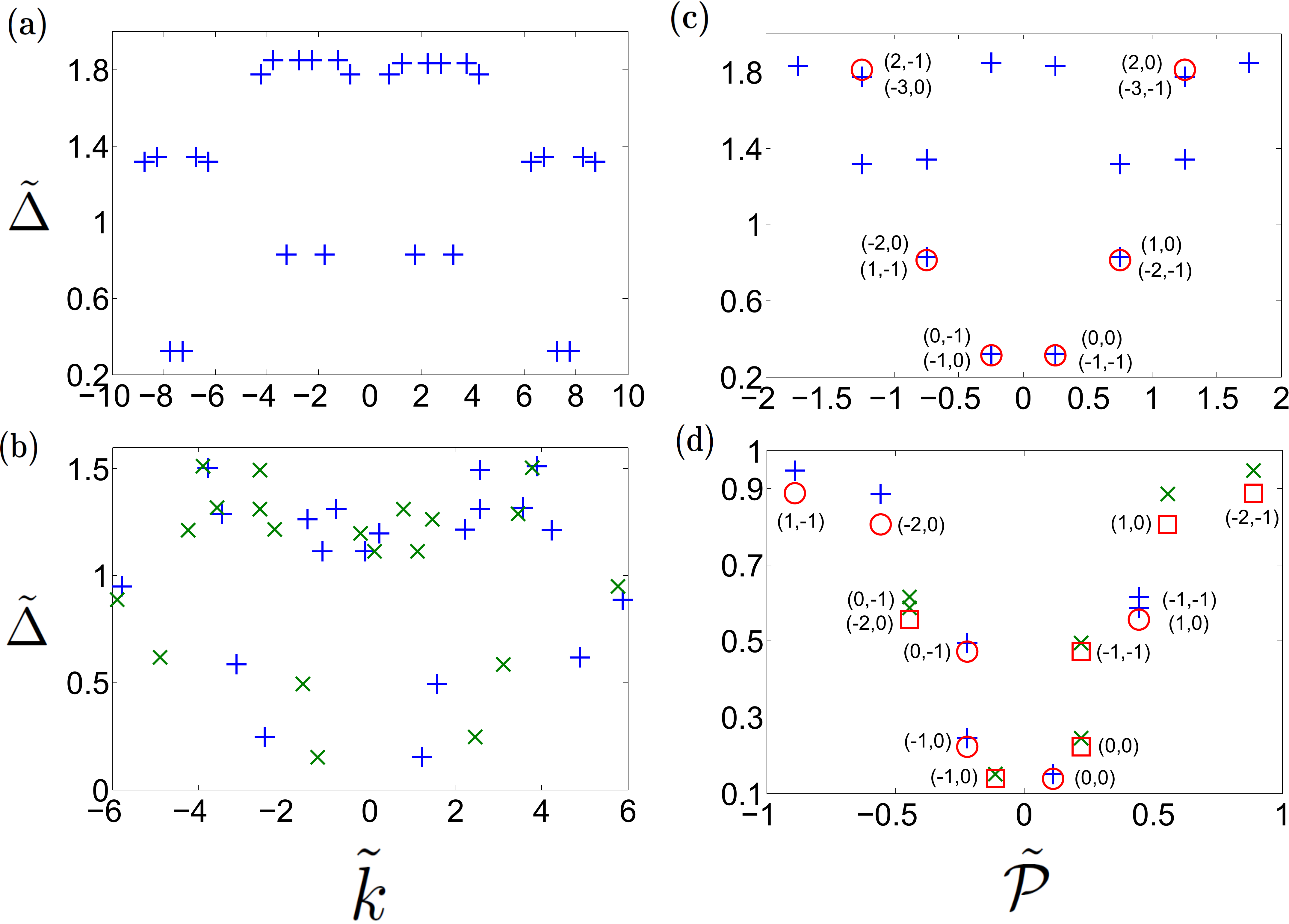}
\caption{
Spectrum of the twisted SPT Hamiltonian 
with respect to the lowest energy $E^{(p)}_{N,0}$
on a ring as a function of the lattice momentum
$\tilde{k}$, with the same values
of 
$
\lambda^{(p)}_{N}
$
as in Fig.~(\ref{fig:spectrum untwisted SPT}).
First few primary states labeled by $(n,m)$.
(a) Spectrum of $\tilde{H}^{(1)}_{2}$ with $M=20$ sites.
(b) Spectrum of $\tilde{H}^{(1)}_{3}$ (+) 
and $\tilde{H}^{(2)}_{3}$ ($\times$) with $M=12$ sites.
(c) Comparison between $\tilde{\Delta}^{(1)}_{2}$ (circles) and numerical results (+)
plotted as a function of the momentum $\tilde{\mathcal{P}}^{(1)}_{2}$. 
All points are two-fold degenerate. Red circles 
represent primary states, while the remaining points
account for descendant states in the CFT spectrum. 
(d) Comparison between $\tilde{\Delta}^{(1)}_{3}$ (circles) and data points (+)
plotted in terms of the momentum $\tilde{\mathcal{P}}^{(1)}_{3}$. 
Same for $\tilde{\Delta}^{(2)}_{3}$ (squares) and data points ($\times$)
plotted in terms of the momentum $\tilde{\mathcal{P}}^{(2)}_{3}$.
}
\label{fig:spectrum of twisted Hamiltonians}
\end{figure}

In Figs.~(\ref{fig:spectrum of twisted Hamiltonians}\,a) 
and~(\ref{fig:spectrum of twisted Hamiltonians}\,b) 
we display the low energy spectrum of the twisted
$\mathbb{Z}_{2}$ and $\mathbb{Z}_{3}$
SPT Hamiltonians with a $\pi$-flux and $2\pi/3$-flux, respectively,
as a function of twisted lattice momentum $\tilde{k}$
defined as 
$
\tilde{T}
=
e^
{
i\,\frac{2\pi}{M}
\tilde{k}
}
$.
The eigenvalues of the primary states 
show very good agreement with 
$\tilde{\Delta}^{(1)}_{2}(n,m;R=2)$
and 
$\tilde{\Delta}^{(1,2)}_{3}(n,m;R=2)$
in Eq.~(\ref{eq:scaling dimensions with flux}), which 
we compare, in Figs.~(\ref{fig:spectrum of twisted Hamiltonians}\,c)
and~(\ref{fig:spectrum of twisted Hamiltonians}\,d),
by folding the spectrum so that the primary states are plotted as a function
of the continuum momenta
$\tilde{\mathcal{P}}^{(1)}_{2}(n,m)$
and
$\tilde{\mathcal{P}}^{(1,2)}_{3}(n,m)$.
Our findings thus establish a 
relationship between the many-body AB effect
both in terms of a long wavelength description in the field theory
as well as in terms of twisted boundary conditions
in a lattice model.

\section{Summary}

We have demonstrated that an intrinsically many-body realization of
the Aharonov-Bohm phenomenon takes place on the edge of a $2$D symmetry-protected many-body system 
in the presence of a background gauge flux.
In our construction we have assumed that edge state is in a gapless
phase and is described by a simple non-chiral Luttinger liquid action with
one right and one left moving propagating modes carrying different 
$\mathbb{Z}_{N}$ charges~\cite{Comment}, in which case, the 
spectrum in the presence of a gauge flux
displays quantization as Eq.(\ref{eq:scaling dimensions with flux})
due to global symmetry protection ($\mathbb{Z}_{N}$ symmetry in our work), 
analogous to the quantization of the energy spectrum of a superconducting
ring due to the $\mathbb{Z}_{2}$ symmetry inherent to superconductors\cite{Byers:1961zz}.
The universal information carried by the counter propagating edge modes 
is that they carry different $\mathbb{Z}_{N}$ charges, which has been 
numerically verified for the $\mathbb{Z}_{2}$ and $\mathbb{Z}_{3}$ SPT classes
in Fig.~(\ref{fig:spectrum untwisted SPT}), 
where this difference is
parametrized by the integer $p \in \{1,...,N-1\}$ that characterizes the SPT class.
This quantum number should remain invariant as long as the SPT order is not destroyed in the
bulk.
The offset in the charges carried by the right and left moving modes has then been shown to reflect itself in the edge spectrum according to Eq.~(\ref{eq:scaling dimensions with flux}) (where $R$ is a non-universal parameter), 
which we have confirmed numerically
in our model Hamiltonians for the $\mathbb{Z}_{2}$ and $\mathbb{Z}_{3}$ SPT classes
in Fig.~(\ref{fig:spectrum of twisted Hamiltonians}).

We have proposed general principles guiding the construction 
of the lattice Hamiltonians,
Eqs.~(\ref{eq:Hamiltonian lattice}) and (\ref{eq:Hamiltonian lattice-twisted a}),
of the bosonic $\mathbb{Z}_{N}$-symmetric
SPT edge states for both the untwisted/twisted (without/with gauge fluxes) cases.

The twisted spectra (i.e. with gauge flux)  characterize all types of $\mathbb{Z}_N$ bosonic anomalies\cite{Wen:2013oza,Wang:2014tia}, which naturally 
serve as ``SPT invariants\cite{Wen:2013ue}'' to detect and distinguish all $\mathbb{Z}_N$ classes of SPT states numerically/experimentally.
(See also recent works\cite{Zaletel,Wang:2014tia}.)

Gauging a non-onsite symmetry of SPT has been noticed relating to Ginsparg-Wilson(G-W) fermion\cite{Ginsparg:1981bj} approach of a lattice field theory problem\cite{Wang:2013yta}.
We remark that our current work achieves gauging a non-onsite 
symmetry for a bosonic system, thus providing an important step towards this direction.
Whether our work can be extended to more general symmetry classes and to fermionic systems (such as U(1) symmetry in G-W fermion approach) is an open question, which we leave for future works. 

 \section*{Acknowledgements} 
We acknowledge useful discussions with 
X. Chen, L. Cincio, D. Gaiotto, T. Senthil, G. Vidal, A. Vishwanath and X.-G. Wen.
We are particularly grateful to Xie Chen on clarifying her works, 
and to Lukasz Cincio for introducing us to some of the numerical methods used here.
%
%
After the Phys.\ Rev.\ B journal publication, JW thanks Yakir Aharonov for his interests in discussing our work in person at Perimeter Institute.
LHS would like to thank the 
ICPT South American Institute for Fundamental Research (ICTP-SAIFR) 
for the hospitality where part of this work was carried out via 
a partnership with the Perimeter Institute.
This research is supported by NSF Grant No.
DMR-1005541, NSFC 11074140, and NSFC 11274192(JW). 
Research at Perimeter Institute is supported by the Government of Canada through Industry Canada and by the Province of Ontario through the Ministry of Economic Development \& Innovation. (LHS and JW)


\appendix

\begin{center}
{\bf Appendix}\\

\end{center}

In Appendix 1, we briefly review the field theory tool for topological states, especially symmetry-protected topological (SPT) states,
but with the emphasis on the canonical quantization, and how 
the global symmetry transformation $\mathcal{S}^{(p)}_{N}$ on the edge is encoded in the canonical quantization.
Using the same formalism, in Appendix 2, we derive the twisted boundary condition due to a gauge flux insertion.

In Appendix 3, we provide our detailed lattice construction (with $\mathbb{Z}_N$ symmetry) for both the untwisted/twisted (without/with gauge flux) cases. 

In Appendix 4, we match each SPT class of our lattice construction to the 3-cocycles in the group cohomology classification.

\section{1. Field Theory Realization of $\mathbb{Z}_N$ SPT States} 

\subsection{1.1. Bulk and boundary actions}
A general framework of categorizing and classifying Abelian topological orders, especially the SPT ones, in 2+1D,
makes use of
Abelian $K$-matrix Chern-Simons theory\cite{Wenbook}.
We now derive the $K$-matrix construction for the SPT order, 
following the pioneering works\cite{LevinGu,Lu:2012dt,Chenggu,Wang:2012am,Hung:2013nla,Ye:2013upa}. 

The intrinsic field theory description of SPT states, on a 2D spatial surface $\mathcal{M}^2$, is the Chern-Simons action
\be
I_{SPT,\mathcal{M}^2}=\frac{1}{4\pi} \int dt\;d^2x K_{IJ}\epsilon^{\mu\nu\rho} a^I_\mu \partial_\nu a^J_\rho\; 
\ee
where $a$ is the intrinsic(or statistical) gauge field, and $K$ is the $K$-matrix which categorizes the SPT orders. 
An SPT state is not intrinsically topologically ordered\cite{Chen:2011pg}, so it has no topological degeneracy\cite{Wenbook,Wang:2012am}.
Ground state degeneracy(GSD) of SPT on the torus is $\GSD=|\det K|=1$\cite{Wenbook,Lu:2012dt,Wang:2012am}, this suggests a constrained canonical form of $K$\cite{Lu:2012dt,Wang:2012am,Ye:2013upa}.

The SPT order is symmetry-protected, so tautologically 
its order is protected by a global symmetry. 
The novel features of SPT distinct from a trivial insulator is its symmetry-protected edge states on the boundary. 
The effective degree of freedom of its 1D edge, $\partial \mathcal{M}^2$, is chiral bosonic field $\phi$, where $\phi$ is
meant to preserve gauge invariance on the bulk-edge under gauge transformation of the field $a$\cite{Wenbook}. The boundary action shows 
\be \label{eq:Ledge}
I_{SPT,\partial \mathcal{M}^2}= \frac{1}{4\pi} \int dt\; dx \; \big( K_{IJ} \partial_t \phi_{I} \partial_x \phi_{J} -V_{IJ}\partial_x \phi_{I}   \partial_x \phi_{J} \big).
\ee

\subsection{1.2 $\mathbb{Z}_N$ symmetry transformation}

The $\mathbb{Z}_N$ symmetry simply requires a rank-2 $K$-matrix, which exhausts all the group cohomology class, $\cH^3(\mathbb{Z}_N,U(1))=\mathbb{Z}_N$,  
\be
K =\bigl( {\begin{smallmatrix} 
0 &1 \\
1 & 0
\end{smallmatrix}}  \bigl).
\ee
The $\mathbb{Z}_N$ symmetry transformation with a $\mathbb{Z}_N$ angle specifies the group element $g$\cite{Lu:2012dt},
\be \label{eq:sym_field}
g_{n}: 
\delta \phi_{g_{n}}= \frac{2\pi}{N} n {\begin{pmatrix} 
1  \\
p  
\end{pmatrix}},\;
\ee
where $p$ labels the $\mathbb{Z}_N$ class of the cohomology group $\cH^3(\mathbb{Z}_N,U(1))=\mathbb{Z}_N$.
Both $n$ and $p$ are module $N$ as elements in $\mathbb{Z}_N$. It can be shown that under
$ \phi_{g_{n}} \to \phi_{g_{n}}+\delta \phi_{g_{n}}$, the action Eq.~(\ref{eq:Ledge}) is invariant, and the $\mathbb{Z}_N$ group structure is realized through $g_{n}^N=\openone$. 
The construction of more general symmetry classes can be found in Ref.\onlinecite{Lu:2012dt, Ye:2013upa}.

\subsection{1.3 Canonical quantization}
Here we go through the canonical quantization of the boson field $\phi_{I}$. 
For canonical quantization, we mean that imposing a commutation relation between $\phi_{I}$ and its conjugate momentum field $\Pi_{I}(x)=\frac{\delta {L}}{\delta (\partial_t \phi_{I} )}=\frac{1}{2\pi} K_{IJ} \partial_x \phi_{J}$. 
Because $\phi_{I}$ is a compact phase of a matter field,
its bosonization contains both zero mode ${\phi_{0}}_{I}$ and winding momentum $P_{\phi_J}$, in addition to non-zero modes\cite{Wang:2012am}: 
\be \label{eq:mode}
\phi_I(x) ={\phi_{0}}_{I}+K^{-1}_{IJ} P_{\phi_J} \frac{2\pi}{L}x+i \sum_{n\neq 0} \frac{1}{n} \alpha_{I,n} e^{-in x \frac{2\pi}{L}}.
\ee
The periodic boundary has size $0\leq x<L$. 
Firstly we impose the commutation relation for zero mode and winding modes, and generalized Kac-Moody algebra for non-zero modes:
\be
[{\phi_{0}}_{I},  P_{\phi_J}]=i\delta_{IJ},\;\; [\alpha_{I,n} , \alpha_{J,m} ]= n K^{-1}_{IJ}\delta_{n,-m}.
\ee
We thus derive canonical quantized fields with the commutation relation:
\bea
[\phi_I(x_1), K_{I'J} \partial_x \phi_{J}(x_2)]&=& {2\pi} i  \delta_{I I'} \delta(x_1-x_2), \label{eq:commutation}\\\;\;
[\phi_I(x_1),\Pi_{J}(x_2)]&=&  i  \delta_{IJ} \delta(x_1-x_2).
\eea

The symmetry transformation of Eq.~(\ref{eq:sym_field}) imples $ \phi_{g_{n}} \to \phi_{g_{n}}+\delta \phi_{g_{n}}$:
\begin{equation}
 {\begin{pmatrix} 
\phi_1 (x)   \\
\phi_2  (x)
\end{pmatrix}} 
\to
 {\begin{pmatrix} 
\phi_1 (x)  \\
\phi_2  (x)
\end{pmatrix}} 
+\frac{2\pi}{N}{\begin{pmatrix} 
1   \\
p  
\end{pmatrix}}.  
\end{equation}
It can be easily checked, using Eq.~(\ref{eq:commutation}), that 
\be \label{eq:globalS}
\mathcal{S}^{(p)}_{N}
=
e^
{
\frac{i}{N}\,
\left(
\int^{L}_{0}\,dx\,\partial_{x}\phi_{2}
+
p\,\int^{L}_{0}\,dx\,\partial_{x}\phi_{1}
\right)
} 
\ee
implements the global symmetry transformation 
\begin{equation}
\mathcal{S}^{(p)}_{N}
 {\begin{pmatrix} 
\phi_1 (x)   \\
\phi_2  (x)
\end{pmatrix}} (\mathcal{S}^{(p)}_{N})^{-1}
=
 {\begin{pmatrix} 
\phi_1 (x)  \\
\phi_2  (x)
\end{pmatrix}} 
+\frac{2\pi}{N}{\begin{pmatrix} 
1   \\
p  
\end{pmatrix}}.  
\end{equation}

\section{2. Twisted boundary condition from a gauge flux insertion}
\label{sec:appendix CS with flux}

Here 
we apply the canonical quantization method to formulate the effect of a gauge flux insertion through a cylinder (an analog of Laughlin thought experiments~\cite{Laughlin1981})
in terms of a twisted boundary condition effect. 
The {\it canonical quantization approach} here can be compared with the alternate {\it path integral approach} motivated in the main text.  
The canonical quantization offers a solid view why the twisted boundary condition resulted from a gauge flux is a quantum effect. We will firstly present the bulk theory viewpoint, then the edge theory viewpoint.

\subsection{2.1 Bulk theory}
Our setting is an external adiabatic gauge flux insertion through a cylinder/annulus. 
Here the gauge field (such as electromagnetic field) couples to (SPT or intrinsic) topologically ordered states, by a coupling charge vector $q_I$.
The bulk term (here we recover the right dimension, while one can set these to be $e=\hbar=c=1$ in the end)
\be
I_{bulk}= 
\int_\mathcal{M}  (c \;dt)\; d^2x\;\Big( (\frac{e^2}{\hbar})\;\frac{K_{IJ}}{4\pi}   \epsilon^{\mu\nu\rho} a^I_{\mu} \partial_\nu a^J_{\rho} +e q^I A_\mu J^\mu_I\Big),
\ee
where $J^\mu_I$ is in a conserved current form
\be
J^\mu_I= (\frac{e }{\hbar}) \frac{1}{2\pi}  \epsilon^{\mu\nu\rho} \partial_\nu a_{\rho,I}.
\ee
From the action, we derive the EOM
\be
J_J^\mu=-q_I \frac{e}{2 \pi} K^{-1}_{IJ} \frac{c}{\hbar}\epsilon^{\mu \nu \rho} \partial_\nu A_\rho.
\ee

From the bulk theory side, an adiabatic flux $\Delta \Phi_B$ induces an electric field ${E}_x$ by Faraday effect, causing a perpendicular current $J_y$ flow to the boundary edge states.
We can explicitly derive the flux effect from Faraday-Maxwell equation in the 2+1D bulk,
\bea 
q_I \Delta \Phi_B&=&-q_I \int dt \int\vec{E} \cdot d\vec{l}=q_I \int dt\;d l_\mu\; c\; \epsilon^{\mu \nu \rho} \partial_\nu A_\rho   \nonumber\\ 
&=&-\frac{2 \pi }{e} K_{IJ}\hbar\int J_{y,J} dt dx =-\frac{2 \pi }{e} K_{IJ}\frac{\hbar}{e} Q_J, \label{bulkQ}
\eea
which relates to the induced charge transported through the bulk, via the Hall effect mechanism. This is a derivation of Laughlin flux insertion argument.
$Q$ is the total charge transported through the bulk, which should condense on the edge of cylinder.

\subsection{2.2 Edge theory}

On the other hand, from the boundary theory side, the induced charge $Q_I$ on the edge can be derived from 
the edge state dynamics affecting winding modes(see Eq.~(\ref{eq:mode})) by 
\be
Q_I=\int J^0_{\partial,I} dx=-\oint^{L}_0 \frac{e}{2\pi}  \partial_x \phi_I dx =-e K^{-1}_{IJ} P_{\phi,J}. \label{boundaryQ}
\ee
Combine Eq.(\ref{bulkQ}) and (\ref{boundaryQ}),
\be
q_I \Delta \Phi_B/(2\pi \frac{\hbar}{e})=\Delta P_{\phi,I}.
\ee

An equivalent interpretation is that the flux insertion twists the boundary conditions of $\phi_I$ field
\bea
\frac{1}{2\pi} (\phi_I (L)-\phi_I (0))&=&\oint^L_0 \frac{1}{2\pi}  \partial_x \phi_I dx = K^{-1}_{IJ} \Delta P_{\phi,J}\;\;\;\;\;\;\\
&=& K^{-1}_{IJ} \;q_J   \Big(\Delta \Phi_B/(2\pi \frac{\hbar}{e}) \Big)
\eea

In the $\mathbb{Z}_N$ symmetry SPT case at hand, we should replace $e$ to the condensate(order parameter) charge $e^*=Ne$.
This affects the unit of $\Delta \Phi_B$ as $2\pi \frac{\hbar}{e^*}$, so $\Delta \Phi_B=2\pi n\frac{\hbar}{Ne}$, and the twisted boundary condition is
\bea
\frac{1}{2\pi} (\phi_I (L)-\phi_I (0))
=K^{-1}_{IJ} \;q_J   \Big( n/N \Big).\;\;\;
\eea

Notice $q_J$ is the crucial coupling in the global symmetry transformation, where we gauge it by minimal coupling to a gauge field $A$ with a term $q^I A_\mu J^\mu_I$.
Here $q_J$ is realized by $(1,p)$ from Eq.(\ref{eq:sym_field}), so
inserting a unit $\mathbb{Z}_N$ flux produces
\be
(\phi_I (L)-\phi_I (0))
=\frac{2\pi}{N}{\begin{pmatrix} 
p   \\
1  
\end{pmatrix}}.   
\ee
In other words, while the global $\mathbb{Z}_N$ symmetry transformation is realized by
\be
\mathcal{S}^{(p)}_{N}
 {\begin{pmatrix} 
\phi_1 (x)   \\
\phi_2  (x)
\end{pmatrix}} (\mathcal{S}^{(p)}_{N})^{-1}
=
 {\begin{pmatrix} 
\phi_1 (x)  \\
\phi_2  (x)
\end{pmatrix}} 
+\frac{2\pi}{N}{\begin{pmatrix} 
1   \\
p  
\end{pmatrix}},  
\ee 
the insertion of a unit $\mathbb{Z}_N$ gauge flux implies the twisted boundary condition
\be \label{eq:twist_BC} 
 {\begin{pmatrix} 
\phi_1 (L)   \\
\phi_2  (L)
\end{pmatrix}} 
=
 {\begin{pmatrix} 
\phi_1 (0)   \\
\phi_2 (0) 
\end{pmatrix}} 
+\frac{2\pi}{N}{\begin{pmatrix} 
p   \\
1  
\end{pmatrix}}.  
\ee

Here $\phi_1(x)$ is realized as the long wavelength description of the rotor angle variable introduced in the maintext, while its conjugate momentum is the angular momentum
\be
L_{\phi_1}(x)=\frac{1}{2\pi} \partial_x \phi_2(x)
,\ee 
where 
\be
[\phi_1(x_1),L_{\phi_1}(x_2)]=i   \delta(x_1-x_2).
\ee

We stress that our result is very different from a seemly similar study in Ref.\onlinecite{Wen:2013ue},
where ``the gauging process'' is 
by coupling the bulk state to an external gauge field $A$, and integrating out the intrinsic field $a$, to get an effective response theory description. However, the twisted boundary condition derived in \cite{Wen:2013ue} 
does not capture the dynamical effect on the edge under gauge flux insertion.
Instead, in our case, we can capture this effect in  Eq.~(\ref{eq:twist_BC}).

\subsection{3. From field theory to lattice model}

Here we motivate the construction of our lattice model from the field theory.
Our lattice model uses the rotor eigenstate $| \phi \rangle$ as basis, where in $\mathbb{Z}_N$ symmetry, $\phi=n (2\pi/N)$, with n is a $\mathbb{Z}_N$ variable.
The conjugate variable of $\phi$ is the angular momentum $L$, which again is a $\mathbb{Z}_N$ variable.
The $| \phi \rangle$ and $| L \rangle$ eigenstates are related by a Fourier transformation, $| \phi \rangle=\sum^{N-1}_{L=0} \frac{1}{\sqrt{N}} e^{i L \phi} | L \rangle$.

\subsection{3.1 General Hamiltonian construction}

The $\mathbb{Z}_N$ class of Hamiltonian may be realized by $H^{(p)}_{N}$, 
with $p\in \mathbb{Z}_N$,
\begin{equation}
\label{eq:Hamiltonian lattice appendix}
\begin{split}
H^{(p)}_{N}
&\,
\equiv
\lambda^{(p)}_{N}
\sum^{M}_{j=1}
h^{(p)}_{N,j}
\\
&\,
=
-
\lambda^{(p)}_{N}
\sum^{M}_{j=1} 
\sum^{N-1}_{\ell = 0}
\left(
S^{(p)}_{N} 
\right)^{-\ell}
( \tau_{j} + \tau^{\dagger}_{j} )  
\left(
S^{(p)}_{N} 
\right)^{\ell},
\end{split}  
\end{equation}  
with the parametrization
\begin{equation}
\tau_j =e^{i 2\pi L_j/N}.
\end{equation}
$S^{(p)}_{N}$ is the $\mathbb{Z}_N$ class of symmetry transformation%

\begin{equation}
\label{eq:symmetry form lattice appendix}
\begin{split}
S^{(p)}_{N}
&\,\equiv
\prod^{M}_{j=1} 
\tau_j  
\prod^{M}_{j=1} 
\exp\Big\{ i \frac{p}{N}  \big[\frac{2\pi}{N}(\delta N_{\text{DW}})_{j,j+1} \big]\Big\}
\\
&\,
\equiv 
\prod^{M}_{j=1} \tau_j \;
\prod^{M}_{j=1}  
e^
{
\frac{i}{N}\,Q^{(p)}_{N}(\sigma^{\dagger}_{j}\sigma_{j+1})
}, 
\end{split}
\end{equation}
where
\begin{equation}
Q^{(p)}_{N}(\sigma^{\dagger}_{j}\sigma_{j+1})
=
\sum^{N-1}_{a=0}\,
q^{(p)}_{N,a}\,(\sigma^{\dagger}_{j}\sigma_{j+1})^{a}.
\end{equation}  
 
Hermiticity of $Q^{(p)}_{N}$ combined with 
$
\sigma^{\dagger}_{j}\sigma_{j+1}
\in
\mathbb{Z}_{N}
$
imply the constraint on the complex coefficients $q_{a}$
(we drop indices $p,N$ to simplify notation and,
in the following, bar denotes complex conjugation):
\begin{equation}
q_{0} 
\in 
\mathbb{R};
~
q_{a}
=
\bar{q}_{N-a},
~
a = 1,...,(N-1)/2
\end{equation}
for odd $N$,
while
\begin{equation}
\begin{split}
q_{0} 
\in 
\mathbb{R};
~
q_{a}
=
\bar{q}_{N-a},
~
a = 1,...,N/2-1;
~
q_{\frac{N}{2}}
\in
\mathbb{R}
\end{split}  
\end{equation}
for even $N$.
The coefficients of the $(N-1)$-th order polynomial
operator 
$
Q^{(p)}_N(\sigma^{\dagger}_{j}\sigma_{j+1})
$
are determined, up to unimportant phases,
by the condition 
\begin{equation}
e^
{
i\,Q^{(p)}_{N}(\sigma^{\dagger}_{j}\sigma_{j+1})
}
=
(\sigma^{\dagger}_{j}\sigma_{j+1})^{p},
\quad
p = 0,...,N-1.
\label{eq:Z_{N} condition on non-onsite term}
\end{equation}  
Solution of Eq.~(\ref{eq:Z_{N} condition on non-onsite term}) 
can be systematically found for each value of $p \in \mathbb{Z}_{N}$ 
giving rise to different symmetry classes. 
Below, for the sake of concreteness, we give explicit forms of the 
symmetry transformations and Hamiltonians for 
$\mathbb{Z}_{2}$ and $\mathbb{Z}_{3}$ groups.

\subsubsection{3.1.1 $\mathbb{Z}_2$ Lattice model}
For $N=2$ lattice model, in the $| \phi \rangle$ basis, we have $| \phi=0 \rangle, | \phi=\pi \rangle$,
and $\omega = e^{i\,\pi} = -1$.

\be
\langle \phi_a | e^{i \phi_j} | \phi_b \rangle  =  {\begin{pmatrix} 
1 & 0  \\
0 & -1  
\end{pmatrix}}_{ab,j}  =\sigma_{ab,j}=(\sigma_{z})_{ab,j} 
\ee

\be
\langle \phi_a | \tau_j  | \phi_b \rangle  =\langle \phi_a |  e^{i2 \pi L_j/N} | \phi_b \rangle =  {\begin{pmatrix} 
0 & 1  \\
1 & 0  
\end{pmatrix}}_{ab,j} =\tau_{ab,j}=(\sigma_{x})_{ab,j}
\ee

The symmetry transformation reads 
\begin{equation}
S^{(p)}_{2}
=\prod^{M}_{j=1}\,\tau_{j}\,
\prod^{M}_{j=1}\,
e^
{
\frac{i}{2}
\,
Q^{(p)}_{2}(\sigma^{z}_{j}\sigma^{z}_{j+1})
},
\end{equation}
where we find, by imposing condition~(\ref{eq:Z_{N} condition on non-onsite term}),
\begin{equation}
Q^{(p)}_{2}(\sigma^{z}_{j}\sigma^{z}_{j+1})
=
p\,
\frac{\pi}{2} 
(
1
-
\sigma^{z}_{j}\sigma^{z}_{j+1}
)
,
~~
p = 0,1.
\end{equation}  
With that, we obtain the Hamiltonian in the trivial class as
\begin{equation}
H^{(0)}_{2}
=
-2\lambda^{(0)}_{2}\,\sum^{M}_{j=1}\,
\sigma^{x}_{j},
\end{equation}  
 and in the non-trivial SPT class as
\begin{equation}
H^{(1)}_{2}
=
-\lambda^{(1)}_{2}\,\sum^{M}_{j=1}\,
\left(
\sigma^{x}_{j} - \sigma^{z}_{j-1}\sigma^{x}_{j}\sigma^{z}_{j+1}
\right).
\end{equation}

\subsubsection{3.1.2 $\mathbb{Z}_3$ Lattice model}

For $N=3$ lattice model, in the $| \phi \rangle$ basis, we have $| \phi=0 \rangle, | \phi=2\pi/3 \rangle, | \phi=4\pi/3 \rangle$,
and $\omega=e^{i 2\pi/3}$,
\be
e^{i \phi_j}  =  {\begin{pmatrix} 
1 & 0 & 0 \\
0 & \omega & 0\\  
0 & 0 & \omega^2
\end{pmatrix}}_j =\sigma_{j} 
\ee
\be
e^{i2 \pi L_j/N} =  {\begin{pmatrix} 
0 & 0 & 1  \\
1 & 0 & 0 \\
0 & 1 & 0  
\end{pmatrix}}_j =\tau_{j} 
\ee

The symmetry transformation reads
\begin{equation}
S^{(p)}_{3}
=
\prod^M_{j=1}\,\tau_{j}\,
\prod^M_{j=1}\,
e^
{
\frac{i}{3}
\,
Q^{(p)}_{3}(\sigma^{\dagger}_{j}\sigma^{}_{j+1})
},
\end{equation}
where we find, by imposing condition Eq.~(\ref{eq:Z_{N} condition on non-onsite term}),
\begin{equation}
\begin{split}
&\,
Q^{(p)}_{3}(\sigma^{\dagger}_{j}\sigma^{}_{j+1})
=
q^{(p)}_{0}
+
q^{(p)}_{1}(\sigma^{\dagger}_{j}\sigma^{}_{j+1})
+
\bar{q}^{(p)}_{1}(\sigma^{\dagger}_{j}\sigma^{}_{j+1})^2
\\
&\,
q^{(p)}_{0}
=
-p\,\frac{2\pi}{3},
~~
q^{(p)}_{1}
=
p\,\frac{\pi}{3}(1 + i/\sqrt{3}),
~~
p = 0,1,2.
\end{split}  
\end{equation}  

With that we obtain the Hamiltonian in the trivial class as
\begin{equation}
H^{(0)}_{3}
=
-3 \lambda^{(0)}_{3}\,\sum^M_{j=1}
(
\tau_{j} + \tau^{\dagger}_{j}
),
\end{equation}  
and in the non-trivial SPT classes $p=1,2$ as
\begin{equation}
\begin{split}
H^{(p)}_{3}
&\,=
- \lambda^{(p)}_{3}
\sum^M_{j=1}\,
\Big\{
\tau_{j}\,
\Big[
\frac{5}{3}
+
\frac{\omega+\bar{\omega}}{3}
\left(
\sigma^{\dagger}_{j-1}\sigma_{j}
+
\sigma_{j-1}\sigma^{\dagger}_{j}
\right)
+
\\
&\,
\Big(
\frac{(1+\omega)}{3}\sigma^{\dagger}_{j}\sigma_{j+1}
+
\frac{2\bar{\omega}}{3}\sigma^{\dagger}_{j-1}\sigma_{j+1}
+
\frac{2\omega}{3}\sigma^{\dagger}_{j-1}\sigma^{\dagger}_{j}\sigma^{\dagger}_{j+1}
\\
&\,+
h.c.
\Big)
\Big]
+
h.c.
\Big\}.
\end{split}
\end{equation}

\subsubsection{3.1.3 $\mathbb{Z}_N$ Lattice model}
For a generic $\mathbb{Z}_N$ lattice model, we have $| \phi=0 \rangle, | \phi=2\pi/N \rangle, \dots, | \phi=2\pi(N-1)/N \rangle$,
and $\omega=e^{i 2\pi/N}$. Apply the Fourier transformation, $| \phi \rangle=\sum^{N-1}_{L=0} \frac{1}{\sqrt{N}} e^{i L \phi} | L \rangle$, in the $| \phi \rangle$ basis, we derive
\be
e^{i \phi_j}  =  {\begin{pmatrix} 
1 & 0 & 0 & 0 \\
0 & \omega & 0 & 0\\  
0 & 0 & \ddots  & 0\\  
0 & 0 & 0 & \omega^{N-1}
\end{pmatrix}}_j =\sigma_{j} 
\ee
\be
  e^{i2 \pi L_j/N} =  {\begin{pmatrix} 
0 & 0 & 0 & \dots &0& 1  \\
1 & 0 & 0 & \dots &0& 0 \\
0 & 1 & 0 & \dots &0& 0 \\
0 & 0 & 1 & \dots &0& 0 \\
\vdots &0 & 0 & \dots &1 & 0 
\end{pmatrix}}_j =\tau_{j} 
\ee

Explicit forms of $S^{(p)}_{N}$ can 
systematically be found by imposing
condition Eq.~(\ref{eq:Z_{N} condition on non-onsite term})
for all $p \in \mathbb{Z}_{N}$. The explicit form of the
symmetry transformation reads
\begin{equation}
S^{(p)}_{N}
=
\prod^{M}_{j=1} 
\tau_j  
\,
\prod^{M}_{j=1}  
e^
{
-i\frac{2\pi}{N^2}p\, 
\Big\{
\left(
\frac{N-1}{2}\,
\right)
\openone
+
\sum^{N-1}_{k=1}\,
\frac
{
(\sigma^{\dagger}_{j}\sigma_{j+1})^k
}
{
(\omega^k - 1)
}
\Big\}
}. 
\end{equation}•

\subsection{3.2 Twisted Boundary Conditions on the lattice model}

We clarify some of the steps leading to an edge Hamiltonian
satisfying twisted boundary conditions accounting for the
presence of one unit of background $\mathbb{Z}_{N}$ gauge flux.
The case with a general number of flux quanta can be equally worked out.

Let $T$ be the lattice translation operator 
satisfying
\begin{equation}
T^{\dagger}\,X_{j}\,T = X_{j+1},
\quad
j = 1,...,M,
\end{equation}  
for any operator $X_{j}$ on a ring such that
$
X_{M+1}
\equiv
X_{1}
$.
It satisfies
$
T^{M}
=
\openone
$.
One can then immediately verify
from Eqs.~(\ref{eq:Hamiltonian lattice appendix})
and~(\ref{eq:symmetry form lattice appendix}) that
$
[S^{(p)}_{N}, T]
=
0
$ 
and
\begin{equation}
T^{\dagger}\,h^{(p)}_{N,j}\,T 
= 
h^{(p)}_{N,j+1},
\end{equation}  
from which follows that 
$
H^{(p)}_{N} 
$
in Eq.~(\ref{eq:Hamiltonian lattice appendix})
is translational invariant, i.e,
\begin{equation}
[H^{(p)}_{N}, T]
=
0.
\end{equation}  

Twisted boundary conditions are implemented 
by defining a modified translation operator
\begin{equation}
\tilde{T}^{(p)}
=
T
\,
e^
{
\frac{i}{N}\,Q^{(p)}_{N}(\sigma^{\dagger}_{M}\sigma_{1})
}
\,
\tau_{1}
\end{equation}  
and seeking a 
twisted Hamiltonian
\begin{equation}
\tilde{H}^{(p)}_{N}
\equiv
\lambda^{(p)}_{N}
\sum^{M}_{j=1}
\tilde{h}^{(p)}_{N,j}
\end{equation}  
under the condition that
\begin{equation}
\left(
\tilde{T}^{(p)}
\right)^{\dagger}
\,
\tilde{h}^{(p)}_{N,j}
\,
\left(
\tilde{T}^{(p)} 
\right)
= 
\tilde{h}^{(p)}_{N,j+1},
\label{eq:translation of tilde h}
\end{equation}  
which then yields
\begin{equation}
[
\tilde{H}^{(p)}_{N}
, 
\tilde{T}^{(p)}
]
=
0.
\label{eq:translation invariance twisted H}
\end{equation}  

We now compute, iteratively, 
$
\left(
\tilde{T}^{(p)}
\right)^{M}
$
[where we use
$
U^{(p)}_{M,1}
=
e^
{
\frac{i}{N}\,Q^{(p)}_{N}(\sigma^{\dagger}_{M}\sigma_{1})
}
$],
\begin{widetext}
\begin{equation}
\begin{split}
&\,
\left(
\tilde{T}^{(p)}
\right)^{2}
=
T
\,
U^{(p)}_{M,1}
\,
\tau_{1}
\,
T
\,
U^{(p)}_{M,1}
\,
\tau_{1}
=
T^{2}
\,
U^{(p)}_{1,2}
\,
\tau_{2}
\,
U^{(p)}_{M,1}
\,
\tau_{1}
\\
&\, \vdots
\\
&\,
\left(
\tilde{T}^{(p)}
\right)^{M}
=
\underbrace
{
T^{M}
}
_
{
= \openone
}
(U^{(p)}_{M-1,M}\tau_{M})
(U^{(p)}_{M-2,M-1}\tau_{M-1})
...
(U^{(p)}_{1,2}\tau_{2})
(U^{(p)}_{M,1}\tau_{1})
\\
&\,
\quad\quad\quad\quad
=
(U^{(p)}_{M-1,M}U^{(p)}_{M-2,M-1}...U^{(p)}_{1,2})
\tau_{M}\,U^{(p)}_{M,1}\,
(\tau_{M-1}\tau_{M-2}...\tau_{1})
\\
&\,
\quad\quad\quad\quad
=
\left(
\prod^{M}_{j=1}U^{(p)}_{j,j+1}
\right)
\left(
U^{(p)}_{M,1}
\right)^{\!\!-1}
\!
\tau_{M}\,U^{(p)}_{M,1}\,\tau^{\dagger}_{M}
\left(
\prod^{M}_{j=1}\tau_{j}
\right)
\\
&\,
\quad\quad\quad\quad
=
\left(
\prod^{M}_{j=1}U^{(p)}_{j,j+1}\,
\right)
\,
e^
{
-\!\frac{i}{N}
Q^{(p)}_{N}(\sigma^{\dagger}_{M}\sigma_{1})
}
e^
{
\frac{i}{N}
Q^{(p)}_{N}(\omega\sigma^{\dagger}_{M}\sigma_{1})
}
\,
\left(
\prod^{M}_{j=1}\tau_{j}
\right).
\end{split}  
\end{equation}  
\end{widetext}
Thus we obtain
\begin{equation}
\tilde{S}^{(p)}_{N}\equiv\left(
\tilde{T}^{(p)}
\right)
^{M}
=
e^
{
\frac{i}{N}\,
\Big[
Q^{(p)}_{N}(\omega\sigma^{\dagger}_{M}\sigma_{1})
-
Q^{(p)}_{N}(\sigma^{\dagger}_{M}\sigma_{1})
\Big]
}
\,
S^{(p)}_{N}.
\end{equation}  

Notice that in trivial case ($p=0$) the relation 
\begin{equation}
\tilde{S}^{(p=0)}_{N}=\left(
\tilde{T}^{(p=0)}
\right)^{M}
=
\prod^{N}_{j=1}\,\tau_{j}
=
S^{(p=0)}_{N}.
\end{equation}  
reduces to to the global 
\textit{onsite} symmetry
$
S^{(p=0)}_{N}
$. 
In this case,
the twisted Hamiltonian commutes with the 
onsite symmetry since
$
0
=
[\tilde{H}^{(p=0)}_{N}, \left(\tilde{T}^{(p=0)}\right)^{M}]
=
[\tilde{H}^{(p=0)}_{N}, S^{(p=0)}_{N}]
$,
and the states in the twisted sector are still labeled by
the global trivial $\mathbb{Z}_{N}$ charges,
corresponding to usual toroidal boundary conditions.
In a non-trivial SPT state ($p \neq 0$), however,
we find
$
0
=
[\tilde{H}^{(p)}_{N}, \left(\tilde{T}^{(p)}\right)^{M}]
\neq
[\tilde{H}^{(p)}_{N}, S^{(p)}_{N}]
$,
so that the twisted Hamiltonian breaks the 
non-trivial
$
\mathbb{Z}_{N}
$
SPT global
symmetry.
We should regard $\left(\tilde{T}^{(p)}\right)^{M}\equiv \tilde{S}^{(p)}_{N}$
as a new {\it twisted symmetry transformation} incorporating the gauge flux effect on the branch cut.

\subsubsection{3.2.1 Twisted boundary conditions for the $\mathbb{Z}_{2}$ SPT state}

We now explicitly work out the twisted Hamiltonian
for the non-trivial $\mathbb{Z}_{2}$ SPT state 
and later mention the general $\mathbb{Z}_{N}$ case.
The global SPT symmetry reads
\begin{equation}
S^{(1)}_{2}
=
\prod^{M}_{j=1}\sigma^{x}_{j}\,
\prod^{M}_{j=1}\,
e^
{
\frac{i}{2}\,Q^{(1)}_{2}(\sigma^{z}_{j}\sigma^{z}_{j+1})
}
=
\prod^{M}_{j=1}\sigma^{x}_{j}\,
\prod^{M}_{j=1}\,
e^
{
\frac{i \pi}{4} [ 1 - \sigma^{z}_{j}\sigma^{z}_{j+1} ]
}
\end{equation}  
Define
$
U_{j,j+1}
\equiv
e^
{
\frac{i \pi}{4} [ 1 - \sigma^{z}_{j}\sigma^{z}_{j+1} ]
}
$.
Then the non-trivial SPT Hamiltonian 
$
H
=
\sum^{M}_{j=1}\,h_{j}
$
(we drop overall constants for simplicity) 
is
\begin{equation}
\begin{split}
&\,h_{j}
=
\sigma^{x}_{j} + S^{-1}\,\sigma^{x}_{j}\,S
\\
&\,=
\sigma^{x}_{j} 
+ 
U^{-1}_{j-1,j}\,U^{-1}_{j,j+1}
\,
\sigma^{x}_{j} 
\,
U_{j-1,j}\,U_{j,j+1}
\\
&\,
=
\sigma^{x}_{j} - \sigma^{z}_{j-1}\sigma^{x}_{j}\sigma^{z}_{j+1},
\end{split}  
\end{equation}  
for
$
j = 1,...,M
$.
The modified translation operator reads
\begin{equation}
\tilde{T}
=
T
\,
U_{M,1}
\,
\sigma^{x}_{1}
=
T
\,
e^
{
\frac{i \pi}{4} [ 1 - \sigma^{z}_{M}\sigma^{z}_{1} ]
}
\,
\sigma^{x}_{1}.
\end{equation}  
We seek a twisted Hamiltonian
$
\tilde{H}
\equiv
\sum^{M}_{j=1} \tilde{h}_{j}
$
that commutes with $\tilde{T}$.
It is a simple exercise to check that
\begin{equation}
\begin{split}
&\,
\tilde{T}^{\dagger}h_{2}\tilde{T}
=
h_{3}
\\
&\,
\tilde{T}^{\dagger}h_{3}\tilde{T}
=
h_{4}
\\
&\,
\vdots
\\
&\,
\tilde{T}^{\dagger}h_{M-2}\tilde{T}
=
h_{M-1}.
\end{split}  
\end{equation}  

We are then led to identify
\begin{equation}
\tilde{h}_{j}
\equiv
h_{j},
\quad
j = 2,...,M-1.
\end{equation}  

We now consider
\begin{equation}
\begin{split}
\tilde{h}_{M}
\equiv
\tilde{T}^{\dagger}h_{M-1}\tilde{T}
=
\sigma^{x}_{1}
\,
U^{-1}_{M,1}
\,
h_{M}
\,
U^{}_{M,1}
\,
\sigma^{x}_{1}
\end{split}  
\end{equation}  
and
\begin{equation}
\begin{split}
\tilde{h}_{1}
\equiv
\tilde{T}^{\dagger}\tilde{h}_{M}\tilde{T}
&\,=
\sigma^{x}_{1}
\, 
U^{-1}_{M,1}
\,
( 
\,
\sigma^{x}_{2}
\, 
U^{-1}_{1,2}
\,
h_{1}
\,
U^{}_{1,2} 
\,
\sigma^{x}_{2}
\,
)
\,
U^{}_{M,1} 
\,
\sigma^{x}_{1}
\\
&\,=
\sigma^{x}_{1} 
\,
\sigma^{x}_{2}
\,
\underbrace
{
\left(
U^{-1}_{M,1}
\,
U^{-1}_{1,2}
\,
h_{1}
\,
U^{}_{1,2} 
\,
U^{}_{M,1} 
\right)
}
_
{
 =\, h_{1}
}
\,
\sigma^{x}_{1}
\,
\sigma^{x}_{2}
\\
&\,
=
\sigma^{x}_{1} 
\,
\sigma^{x}_{2}
\,
h_{1}
\,
\sigma^{x}_{1}
\,
\sigma^{x}_{2}.
\end{split}  
\end{equation}  
Now it remains to be shown that
$
\tilde{T}^{\dagger}\tilde{h}_{1}\tilde{T}
=
\tilde{h}_{2}
=
h_{2}
$.
And
indeed
\begin{equation}
\begin{split}
\tilde{T}^{\dagger}\tilde{h}_{1}\tilde{T}
&\,=
\sigma^{x}_{1} 
\,
U^{-1}_{M,1}
\,
( 
\,
\sigma^{x}_{2} 
\,
\sigma^{x}_{3}
\,
h_{2}
\,
\sigma^{x}_{2}
\,
\sigma^{x}_{3}
\,
)
\,
U^{}_{M,1}
\,
\sigma^{x}_{1}
\\
&\,=
\sigma^{x}_{1}
\,
\sigma^{x}_{2} 
\,
\sigma^{x}_{3}
\,
h_{2}
\,
\sigma^{x}_{1}
\,
\sigma^{x}_{2}
\,
\sigma^{x}_{3}
\\
&\,=
h_{2}.
\end{split}  
\end{equation}  
So we have found new terms $\tilde{h}_{j}$
such that 
$
\tilde{T}^{\dagger}\tilde{h}_{j}\tilde{T}
=
\tilde{h}_{j+1}
$,
thus implying that
$
[\tilde{T}, \tilde{H} ] = 0
$.

Explicitly, the twisted Hamiltonian 
for $\mathbb{Z}_{2}$ non-trivial SPT state reads
\begin{subequations}
\begin{equation}
\tilde{H}
=
\sum^{M}_{j=1} \tilde{h}_{j}
\end{equation}  
where
\begin{equation}
\begin{split}
&\,
\tilde{h}_{1}
=
\sigma^{x}_{1} 
\,
\sigma^{x}_{2}
\,
h_{1}
\,
\sigma^{x}_{1}
\,
\sigma^{x}_{2}
=
\sigma^{x}_{1}
+
\sigma^{z}_{M}
\,\sigma^{x}_{1}
\,\sigma^{z}_{2}
\\
&\,
\tilde{h}_{2}
=
h_{2}
=
\sigma^{x}_{2}
-
\sigma^{z}_{1}
\,
\sigma^{x}_{2}
\,
\sigma^{z}_{3}
\\
&\,
\vdots
\\
&\,
\tilde{h}_{M-1}
=
h_{M-1}
=
\sigma^{x}_{M-1}
-
\sigma^{z}_{M-2}
\,
\sigma^{x}_{M-1}
\,
\sigma^{z}_{M}
\\
&\,
\tilde{h}_{M}
=
\sigma^{x}_{1}
\,
U^{-1}_{M,1}
\,
h_{M}
\,
U^{}_{M,1}
\,
\sigma^{x}_{1}
=
\sigma^{y}_{M}
\,
\sigma^{z}_{1}
+
\sigma^{z}_{M-1}
\,
\sigma^{y}_{M}.
\end{split}  
\end{equation}  
\end{subequations}  

\subsubsection{3.2.2 Twisted boundary conditions for the $\mathbb{Z}_{N}$ SPT state}

Generalization to the $\mathbb{Z}_{N}$ case
follows very similar lines as the $\mathbb{Z}_{2}$
case above. We have for the twisted Hamiltonian
(again we drop overall constants)
\begin{subequations}
\begin{equation}
\tilde{H}^{(p)}_{N}
=
\sum^{M}_{j=1} \tilde{h}^{(p)}_{N,j}
\end{equation}  
where
\begin{equation}
\begin{split}
&\,
\tilde{h}^{(p)}_{N,1}
=
\tau^{\dagger}_{1}
\,
\tau^{\dagger}_{2}
\,
h^{(p)}_{N,1}
\,
\tau^{}_{1}
\,
\tau^{}_{2}
\\
&\,
\tilde{h}^{(p)}_{N,2}
=
h^{(p)}_{N,2}
\\
&\,
\vdots
\\
&\,
\tilde{h}^{(p)}_{N,M-1}
=
h^{(p)}_{N,M-1}
\\
&\,
\tilde{h}_{N,M}
=
\tau^{\dagger}_{1}
\,
\Big(U^{(p)}_{M,1}\Big)^{-1}
\,
h^{(p)}_{N,M}
\,
U^{(p)}_{M,1}
\,
\tau_{1},
\end{split}  
\end{equation}  
\end{subequations}  
where
$
U^{(p)}_{M,1}
=
e^
{
\frac{i}{N}
\,
Q^{(p)}_{N}
(
\sigma^{\dagger}_{M}\sigma_{1}
)
}
$.
One can easily verify that 
Eqs.~(\ref{eq:translation of tilde h}) and~(\ref{eq:translation invariance twisted H})
are satisfied.

\subsection{4. Correspondence in Group Cohomology and non-trivial 3-cocycles from MPS projective representation}

Here we map our lattice construction to the 3-cocycles in the group cohomology classification for each SPT class.
Importantly, we notice that the non-onsite piece in $S^{(p)}_N$ is 
\bea
U_{j,j+1} &\equiv& e^
{
i\,Q^{(p)}_{N}(\sigma^{\dagger}_{j}\sigma_{j+1})
}
 =\exp[ \frac{i}{N} \sum^{N-1}_{a=0}\,
q_{a}\,(\sigma^{\dagger}_{j}\sigma_{j+1})^{a} ] \;\;\;\;\;\;\;\\
&\equiv&\exp\Big\{ i \frac{p}{N}  \big[\frac{2\pi}{N}(\delta N_{\text{DW}})_{j,j+1} \big]\Big\} 
\eea
We seek a quantum rotor description of the above form. We claim that 
\be \label{eq:Urotor}
U_{j,j+1} 
=\exp[ i \frac{p}{N}  (\phi_{1,j+1} - \phi_{1,j})_r],   
\ee
which is equivalent to (i) the domain wall picture using rotor angle variables  
(here $(\phi_{1,j+1} - \phi_{1,j})_r$, where subindex $r$ means that we take the module $2\pi$ on the angle~\cite{Chen:2012hc}), and to (ii) the field theory formalism in Eq.~(\ref{eq:globalS}). 

The reason follows: as we mention in the $p$-th case of $\mathbb{Z}_N$ class, we
impose the constraint
\be
U_{j,j+1}^N=(\sigma^{\dagger}_{j}\sigma_{j+1})^p
\ee
to solve the polynomial ansatz $\sum^{N-1}_{a=0}\,
q_{a}\,(\sigma^{\dagger}_{j}\sigma_{j+1})^{a}$.
This is equivalent to the fact that 
\bea
U_{j,j+1}^N &=&(\sigma^{\dagger}_{j}\sigma_{j+1})^p=(\exp[ i  \phi_{1,j}]^\dagger \exp[ i  \phi_{1,j+1}] )^p\;\;\;\\
&=&\exp[ i  p (\phi_{1,j+1} - \phi_{1,j})_r],
\eea
since $\exp[ i  \phi_{1,j}]_{ab}= \langle \phi_a | e^{i \phi_j} | \phi_b \rangle = \sigma_{ab,j}$.
Therefore, 
the domain wall variable $(\delta N_{\text{DW}})_{j,j+1}$ indeed counts the number of units of $\mathbb{Z}_N$ angle between sites $j$ and $j+1$, so
$(2\pi/N)(\delta N_{\text{DW}})_{j,j+1}$$=\phi_{1,j+1} - \phi_{1,j}$.
We thus have shown Eq.~(\ref{eq:Urotor}), and have confirmed affirmatively that 
our approach of lattice regularization is indeed a rotor realization in Ref.\onlinecite{Chen:2012hc} with the same symmetry transformation $S^{(p)}_N$, but captures much more
 than the low energy rotor model there.

The argument on non-trivial 3-cocycles from matrix product states(MPS) projective representation follows closely to Ref.\onlinecite{Chen:2012hc}.
We start from writing the symmetry transformation $S^{(p)}_N$ in terms of the rotor variable,
this is achieved based on the mapping derived above. So 
\be
S^{(p)}_N\equiv \prod^{M}_{j=1}\tau_{j}\,
\prod^{M}_{j=1}U^{(p)}_{j,j+1}=\prod_j  e^{i 2\pi L_j/N} \cdot \exp[ i \frac{p}{N}  ( \phi_{1,j+1}-\phi_{1,j} )_r
].
\ee
We then formulate $S^{(p)}_N$ as the MPS with the form: 
\be
S^{(p)}_N=\sum_{ \{j,j'\}} \tr[T^{j_1 j_1'}_{\alpha_1 \alpha_2}  T^{j_2 j_2'}_{\alpha_2 \alpha_3} \dots T^{j_M j_M'}_{\alpha_M \alpha_1}] |   j'_1, \dots, j'_M\rangle \langle j_1, \dots, j_M |.
\ee
Here $j_1,j_2, \dots, j_M$ and $j'_1,j'_2, \dots, j'_M$ are labeled by input/output physical eigenvalues (here $\mathbb{Z}_N$ angle), the subindices $1,2,\dots, M$ are the physical site indices.
There are also inner indices $\alpha_1, \alpha_2, \dots, \alpha_M$ which are traced in the end.
Summing over all the operation from $\{j,j'\}$ indices is supposed to reproduce the symmetry transformation operator $S^{(p)}_N$.
This tensor $T$ is suggested~\cite{Chen:2012hc} to be, (with the $\mathbb{Z}_N$ angle element $\frac{2\pi k}{N}$)
\bea
&&(T^{\phi_{in}, \phi_{out}})^{(p)}_{\varphi_\alpha,\varphi_\beta,N}( \frac{2\pi k}{N} )=\delta(\phi_{out}-\phi_{in}-\frac{2\pi}{N}k) \nonumber\\
&&\;\;\;\;\;\cdot \int d\varphi_\alpha d\varphi_\beta | \varphi_\beta \rangle \langle \varphi_\alpha | \delta(\varphi_\beta-\phi_{in}) e^{ipk(\varphi_\alpha-\phi_{in})_r
/N}\;\;\;\;\;\;
\eea

\begin{widetext}
We verify the tensor $T$ by computing $S^{(p)}_N$ ,
\bea
&&S^{(p)}_N=\sum_{ \{j,j'\}} \tr[T^{\phi^1_{in}, \phi^1_{out}}_{\varphi_{\alpha_1} \varphi_{\alpha_2}}  T^{\phi^2_{in}, \phi^2_{out}}_{\varphi_{\alpha_2} \varphi_{\alpha_3}} \dots T^{\phi^M_{in}, \phi^M_{out}}_{\varphi_{\alpha_M} \varphi_{\alpha_1}}] |   \phi^1_{out}, \phi^2_{out}, \dots, \phi^M_{out}\rangle \langle \phi^1_{in}, \phi^2_{in}, \dots, \phi^M_{in} |\\
&&=e^{ i \frac{p}{N} \big( (\phi^2_{in} - \phi^1_{in})_r +(\phi^3_{in} - \phi^2_{in})_r+\dots + (\phi^1_{in} - \phi^M_{in})_r\big)}
 |   \phi^1_{in}+\frac{2\pi}{N}, \phi^2_{in}+\frac{2\pi}{N}, \dots, \phi^M_{in}+\frac{2\pi}{N}\rangle \langle \phi^1_{in}, \phi^2_{in}, \dots, \phi^M_{in} |\\
&&=e^{ i \frac{p}{N} \big( \sum^{M}_{j=1} (\phi^{j+1}_{in} - \phi^j_{in})_r\big)}  
|\dots, \phi^j_{in}+\frac{2\pi}{N}, \dots\rangle \langle \dots, \phi^j_{in}, \dots|\\
&&= {\prod_j} \exp[ i \frac{p}{N}  (\phi_{1,j+1} - \phi_{1,j})_r ] \cdot \prod_j  e^{i 2\pi L_j/N},
\eea
\end{widetext}
which justifies the claim for MPS of $S^{(p)}_N$.

To find out the projective representation $e^{i \theta(g_1,g_2, g_3)}$ of this tensors $T(g_1), T(g_2), T(g_3)$ acting on three neighbored sites, we follow the fact that
\be
P_{g_1,g_2}^\dagger T(g_1) T(g_2)P_{g_1,g_2}= T(g_1\cdot g_2)
\ee
and contracting the three neighbored-site tensors in two different orders,
\be \label{eq:projective-3cocycle}
(P_{g_1,g_2} \otimes I_3) P_{g_1g_2, g_3} \simeq e^{i \theta(g_1,g_2, g_3)} ( I_1 \otimes P_{g_2,g_3} ) P_{g_1, g_2g_3}.
\ee
Here $\simeq$ means the equivalence is up to a projection out of un-parallel state transformation.

To derive $P_{g_1,g_2}$, notice that $P_{g_1,g_2}$ inputs one state and output two states. This has the expected form,
\bea
P^{(p)}_{N,m_1,m_2}=\int d\phi_{in} | \phi_{in}+\frac{2\pi}{N}m_2 \rangle |  \phi_{in} \rangle \langle \phi_{in} | \nonumber \\
\cdot \; e^{-ip\phi_{in}[m_1+m_2-(m_1+m_2)_N] /N},
\eea
where $(m_1+m_2)_N$ with subindex $N$ means taking the value module $N$.

In order to derive $\theta(g_1,g_2, g_3)$, we start by contracting $T^{(p)}_N(m_1)$ and $T^{(p)}_N(m_2)$ firstly, and then the combined tensor contracts with $T^{(p)}_N(m_3)$ gives:
\bea \label{eq:projective-1}
&&(P_{g_1,g_2} \otimes I_3) P_{g_1g_2,g_3}  \nonumber \\
&&=\int d\phi_{in}     |\phi_{in}+\frac{2\pi}{N}(m_2+m_3) \rangle  |\phi_{in}+\frac{2\pi}{N}m_3 \rangle  | \phi_{in} \rangle \langle \phi_{in} | \nonumber\\
&& \cdot \;e^{-i p \phi_{in}(m_1+m_2+m_3-(m_1+m_2+m_3)_N)} \nonumber\\ 
&& \cdot \; e^{-i p\frac{2\pi}{N} m_3\frac{m_1+m_2-(m_1+m_2)_N}{N}},\;\;
\eea
which form inputs one state $\langle \phi_{in} |$ and outputs three states $ |\phi_{in}+\frac{2\pi}{N}(m_2+m_3) \rangle$, $|\phi_{in}+\frac{2\pi}{N}m_3 \rangle$ and  $| \phi_{in} \rangle$.

On the other hand, one can contract $T^{(p)}_N(m_2)$ and $T^{(p)}_N(m_3)$ firstly, and then the combined tensor contracted with $T^{(p)}_N(m_1)$ gives:
\bea \label{eq:projective-2}
&&( I_1 \otimes P_{g_2,g_3} ) P_{g_1, g_2g_3}  \nonumber \\
&&=\int d\phi_{in}     |\phi_{in}+\frac{2\pi}{N}(m_2+m_3) \rangle  |\phi_{in}+\frac{2\pi}{N}m_3 \rangle  | \phi_{in} \rangle \langle \phi_{in} | \nonumber\\
&& \cdot \;e^{-i p \phi_{in}(m_1+m_2+m_3-(m_1+m_2+m_3)_N)},  
\eea
again which form inputs one state $\langle \phi_{in} |$ and outputs three states
$ |\phi_{in}+\frac{2\pi}{N}(m_2+m_3) \rangle$, $|\phi_{in}+\frac{2\pi}{N}m_3 \rangle$ and $| \phi_{in} \rangle$. 
From Eq.(\ref{eq:projective-3cocycle}),(\ref{eq:projective-1}),(\ref{eq:projective-2}), we derive:
\be
e^{i \theta(g_1,g_2, g_3)} =e^{-i p\frac{2\pi}{N} m_3\frac{m_1+m_2-(m_1+m_2)_N}{N}},
\ee
which indeed is the 3-cocycle in the third cohomology group $\cH^3(\mathbb{Z}_N,U(1))=\mathbb{Z}_N$. We thus verify that the projective representation $e^{i \theta(g_1,g_2, g_3)}$ from MPS tensors corresponds to 
the group cohomology approach\cite{Chen:2011pg}. This demonstrates that our lattice model construction completely maps to all classes of SPT, as we aimed for.

\end{document}